\documentclass[onecolumn,3p]{elsarticle}

\usepackage[utf8]{inputenc}
\usepackage[english]{babel}
\usepackage{amsmath}
\usepackage{natbib}
\setcitestyle{authoryear,round,semicolon}

\biboptions{sort&compress}

\usepackage{siunitx}
\usepackage{placeins} 
\usepackage{subfig}
\usepackage{graphicx}
\usepackage{amssymb}
\usepackage{multicol}
\usepackage{wrapfig}
\usepackage{cancel}
\usepackage{multirow}
\usepackage{hyperref}
\DeclareGraphicsExtensions{.png, .jpg, .jpeg, .pdf, .gif, .svg}
\usepackage[dvipsnames,table,xcdraw]{xcolor}

\newcommand{\tild}{~}
\newcommand{\equationame}{Eq.\tild}

\newcommand{\fig}{Fig.~}
\newcommand{\figs}{Fig.s~}
\newcommand{\tab}{Tab.\tild}
\newcommand{\sectionname}{Sec.\tild}
\newcommand{\millm}{\, \si{mm}}
\newcommand{\mum}{\, \si{\mu m}}
\newcommand{\megap}{\, \si{MPa}}
\newcommand{\gigap}{\, \si{GPa}}
\newcommand{\dV}{\,\text{dV}}
\newcommand{\dA}{\,\text{dA}}
\newcommand{\dr}{\,\text{dr}}

\newcommand{\Gc}{G_\mathrm{c}}
\newcommand{\sigmac}{\sigma_\mathrm{c}}

\newcommand{\bsigma}{\boldsymbol{\sigma}}
\newcommand{\bepsilon}{\boldsymbol{\epsilon}}
\newcommand{\bu}{\mathbf{u}}
\newcommand{\bg}{\mathbf{g}}
\newcommand{\bn}{\mathbf{n}}
\newcommand{\bt}{\mathbf{t}}
\newcommand{\bN}{\mathbf{N}}
\newcommand{\bL}{\mathbf{L}}
\newcommand{\bR}{\mathbf{R}}
\newcommand{\bx}{\mathbf{x}}
\newcommand{\bp}{\mathbf{p}}
\newcommand{\bB}{\mathbf{B}}

\journal{Journal of the Mechanics and Physics of Solids}

\begin{document}

\begin{frontmatter}
 
\title{A coupled approach to predict cone-cracks in spherical indentation tests with smooth or rough indenters}

\author[IMT]{M.R.~Marulli}
\author[IMT]{J.\tild Bonari}
\author[US]{J.~Reinoso}
\author[IMT]{M.~Paggi\corref{mycorrespondingauthor}}
\cortext[mycorrespondingauthor]{Corresponding author}
\ead{marco.paggi@imtlucca.it}
\address[IMT]{IMT School for Advanced Studies Lucca, Piazza San Francesco 19, 55100 Lucca, Italy}
\address[US]{Grupo de Elasticidad y Resistencia de Materiales, Escuela Técnica Superior de Ingeniería, Universidad de Sevilla, Camino de Los Descubrimientos s/n, 41092 Sevilla, Spain}

\begin{abstract}
Indentation tests are largely exploited in experiments to characterize the mechanical and fracture properties of the materials from the resulting crack patterns. This work proposes an efficient theoretical and computational framework, \textcolor{black}{whose implementation is detailed for 2D axisymmetric and 3D geometries}, to simulate indentation-induced cracking phenomena caused by non-conforming contacts with indenter profiles of arbitrary shape. The formulation hinges on the coupling of the MPJR (eMbedded Profile for Joint Roughness) interface finite elements \textcolor{black}{which embed the indenter profile to solve the contact problem between non-planar bodies efficiently} and the phase-field for brittle fracture to simulate crack evolution and nonlocal damage in the substrate. The novel framework is applied to predict cone-crack formation in the case of indentation tests with smooth spherical indenters, with validation against experimental data. Then, the methodology is employed for the very first time in the literature to assess the effect of surface roughness superimposed on the shape of the smooth spherical indenter. \textcolor{black}{In terms of physical insights, numerical predictions quantify the dependencies of the critical load for crack nucleation and the crack radius on the amplitude of roughness in comparison with the behavior of smooth indenters. Again, the consistency with available experimental trends is noticed}.

This article has been published on the Journal of the Mechanics and Physics of Solids at \url{https://doi.org/10.1016/j.jmps.2023.105345}
\end{abstract}


\end{frontmatter}

\section{Introduction}

An indentation test consists of a sharp or spherical hard tip pressed onto a sample to be mechanically characterized. When the tip is removed after indentation, an impression is left on the target material, and the information on the resulting inelastic deformation can be exploited to assess the mechanical properties of the substrate, see \citep{Hutchings} for a review of the development of the method before and after 1950. For instance, the method has been largely applied to estimate the hardness $H$ of an elasto-plastic material, being $H$ the material resistance to plastic deformation, defined as the indentation load divided by the area of impression: $H = P_m/A_C$. In the above equation, $A_C$ is the projected contact area between the indenter tip and the surface evaluated at the indentation load $P_m$. 

In case of brittle and quasi-brittle materials, like glass, silicon, ceramics, etc., cracks develop during indentation. 
The dimension of these cracks and the indentation loads are related to other material properties of the specimen, and in particular, to the fracture toughness \citep{Lawn1998}.

Different types of indentation tests can be performed according to the tip's dimension and shape. Sharp pyramidal indenters are used in Vickers and Berkovich indentation tests, while the Brinell indentation exploits a spherical tip. For this reason, indentation tests can be applied to different classes of materials, from ceramics and metals, to composites and bio-materials, and at different scales. In this regard, the crack morphology generally depends on the indenter shape: median, radial, and lateral cracks can be recognized in the post-indentation samples \citep{Johanns2014, Schneider2012}. 

\begin{figure}[h]
    \centering
    \subfloat[]{
    \includegraphics[width=0.25\textwidth]{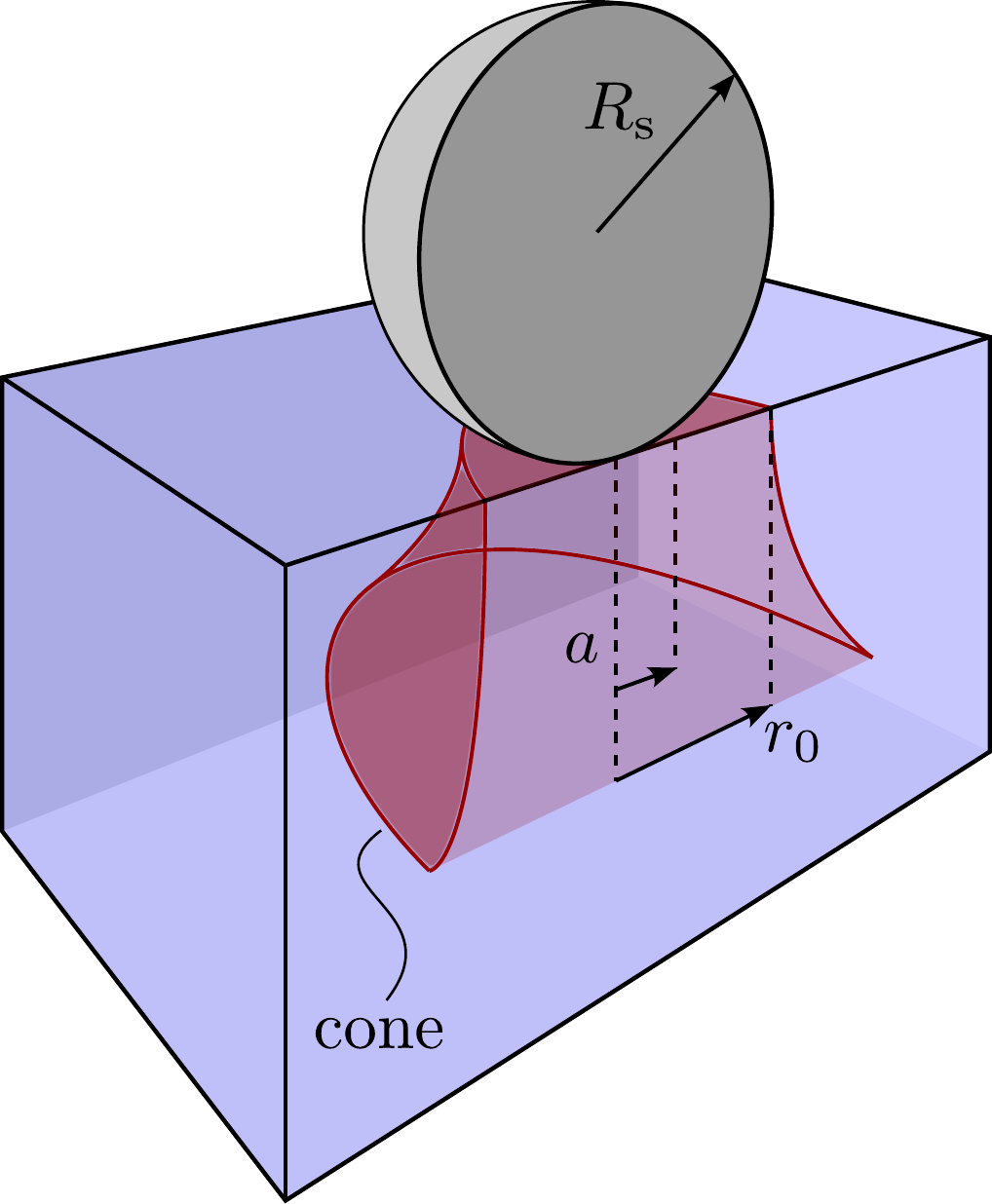}} \qquad \qquad \qquad
    \subfloat[]{
    \includegraphics[width=0.40\linewidth]{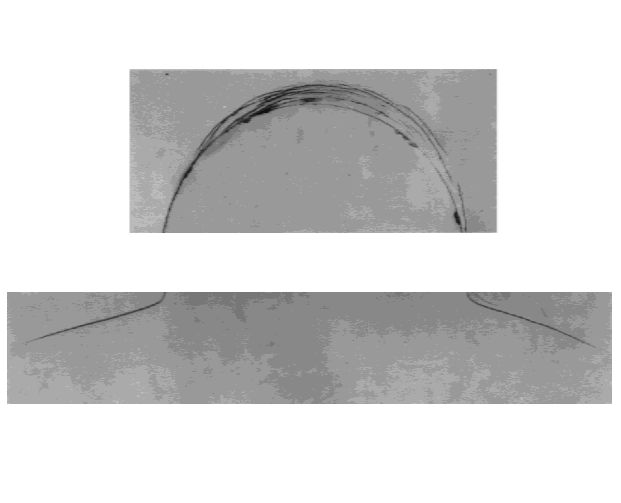}}
    \caption{Hertzian fracture topology: (a) sketch of a spherical indentation test and cone-shaped crack; (b) top view and cross-section of the Hertzian indentation crack in a Silicon Nitride specimen caused by a Tungsten Carbide sphere with $R_s=1.98\millm$, adapted from \citep{Lawn1998}.}
    \label{fig:cone}
\end{figure}

Spherical indentation tests on quasi-brittle materials (see the sketch in \fig\ref{fig:cone}) present the complexity related to the fact that the contact is non-conformal, i.e., the contact area between the sphere and the substrate grows with the applied load. Fracture is characterized by the occurrence of cone-shaped cracks. Moreover, by further increasing the load level, the expanding contact circle may reach and overcome the position of the surface ring crack, resulting in the generation of secondary ring cracks \citep{Lawn1998}.
The formation of such cracks has been pioneeringly studied by \cite{hertz1882} at the end of the 19th Century. By assuming frictionless contact between two elastic solids, the Hertzian description of the stress field under a spherical indenter was used to interpret the fracture nucleation and propagation stages. The fracture pattern develops under the form of a surface ring crack close to the contact boundaries, which then propagates as a cone-crack in the substrate by further increasing the applied load, see the sketch in \fig\ref{fig:cone}. Most of the early experiments on cone fracture were conducted on glass, notably soda–lime glass, or PMMA, under static loading or by considering a free-fall of a spherical indenter onto the specimen. The transparency of these brittle materials allows easy visualization of the crack growth \citep{Puttick1978, Ritter1988, Lawn1998, Schneider2012, Lee2012}. 

\begin{figure}[h]
    \centering
    \subfloat[]{
    \includegraphics[width=0.25\textwidth]{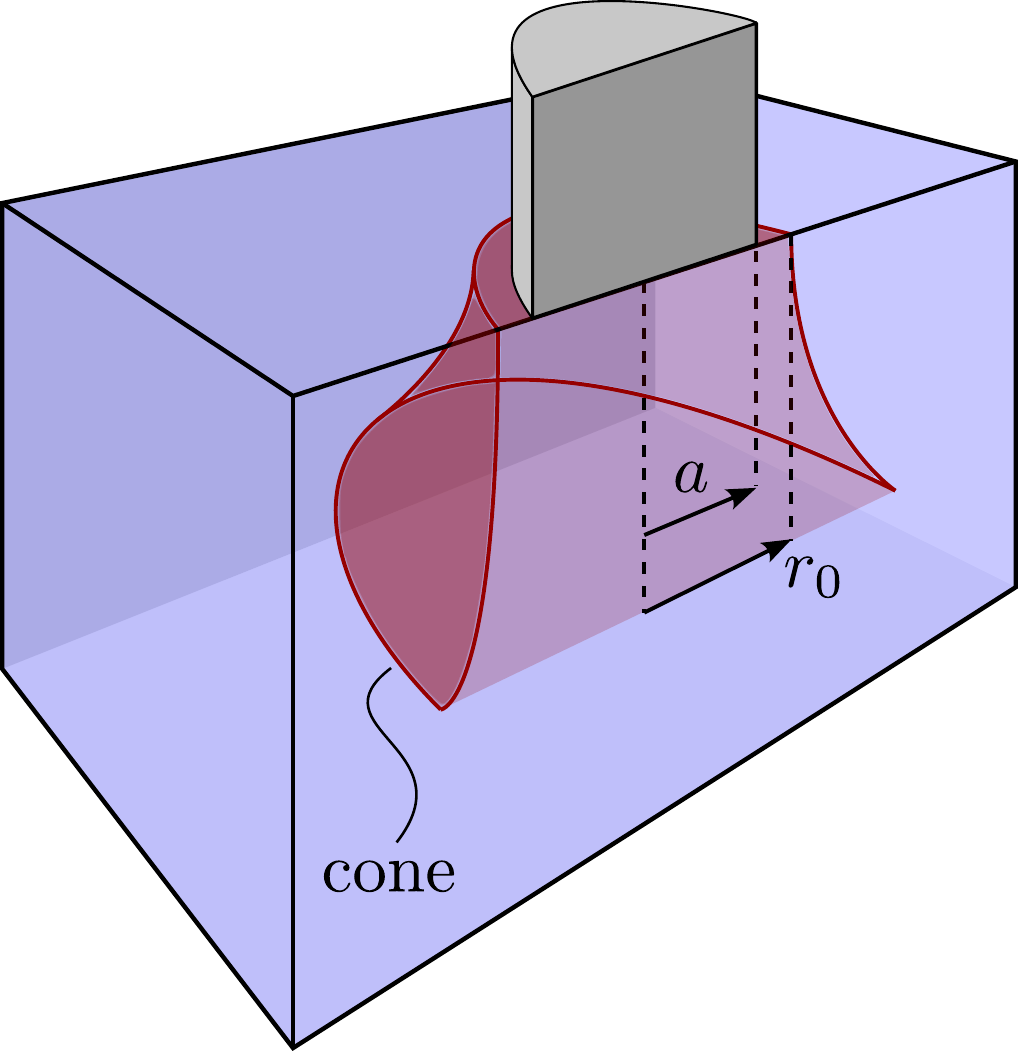}}
    \qquad \qquad \qquad
    \subfloat[]{
    \includegraphics[width=0.45\linewidth]{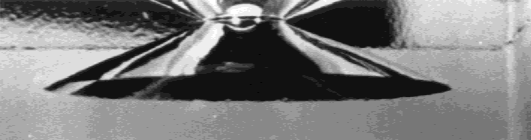}}
    \caption{Indentation fracture caused by a flat-ended cylinder: (a) sketch of the indentation test and crack topology; (b) cone-shaped crack obtained with a cylindrical punch in soda-lime glass, adapted from \citep{Lawn1998}.}
    \label{fig:soda}
\end{figure}

A cone-crack topology can also be obtained by indenting a substrate with a cylindrical flat punch, see \fig\ref{fig:soda}. In this case, the theoretical analysis of the flat cylinder problem is easier than the spherical one since the type of contact is conformal, i.e., the extension of the contact area does not change with the applied load, and, in this case, it is given by the punch base area. 

An experimental comparison between the flat punch and the spherical indenter geometries has been conducted in \citep{Mouginot1985}. For both indenting geometries, the experimental tests showed that the ring-crack radius $r_0$ is always larger than the contact radius at the crack onset $a_c$: the ratio $r_0/a_c$ varies from $1.1$ to $1.4$ for the spherical indentation tests, and it is even higher for the flat punch. Moreover, the ratio $r_0/a_c$ was found to be a decreasing function of the indenter radius $R_s$ \citep{Mouginot1985, Conrad1979}.

After Hertz, \cite{auerbach1893} extensively studied the effect of the indenter size of spherical indenters on cone-crack initiation and gave an empirical linear relationship between the critical indentation load $P_c$ and the sphere radius ($P_c \propto R_s$, known as Auerbach's law). Later, it was noticed that the linear relationship holds only for small indenters when $a$ is comparable with $R_s$. The linear dependency has also been questioned since experimental observations in \citep{Tillett1956} showed that $P_c$ asymptotically grows with $R_s^2$ for larger radii. 

As an advancement from stress theories that have been applied to
assess the critical load for damage nucleation and to predict the direction of crack growth based on the principle stress direction, see e.g. \citep{Mouginot1985} for flat punches, the application of linear elastic fracture mechanics (LEFM) has been fostered to study more in detail the problem of crack growth. In \citep{Tillett1956, Roesler1956, Kocer1998}, LEFM has been used to simulate the propagation of a crack from an existing flaw placed on the surface and to assess the evolution of the crack inclination based on the updated stress field during crack growth, which is clearly different from the estimate based on the uncracked configuration on the base of stress theories. For more details on the experimental and theoretical analyses of the indentation-induced fracture, see, e.g., the review articles in \citep{Kocer2003,guin2019}.


As far as numerical approaches are concerned, an incremental finite element model has been employed in \citep{Kocer2003} to determine the trajectories of the Hertzian cone cracks for various initiation radii by evaluating the change of the stress field caused by crack propagation. The preferred direction of crack growth for its incremental extension was determined using the criterion of maximum strain energy release rate \citep{SUN2012}.  

Cone cracking resulting from spherical indentation has also been simulated using the extended finite element method (XFEM) implemented in Abaqus in \citep{Marimuthu2017} to improve remeshing operation due to crack growth.
A more recent analysis can be found in \citep{Strobl2017}, where the Hertzian cone crack obtained with a cylindrical punch has been analyzed in the context of Finite Fracture Mechanics (FFM). This approach requires the simultaneous fulfillment of a strength and a fracture criterion that involve two well-defined material parameters, i.e., the (tensile) strength $\sigmac$ and the specific (i.e. per unit area) fracture energy $G_c$ \citep{Leguillon2002, Leguillon2018, DOITRAND2022}. The FFM approach in \citep{Strobl2017} reproduced the decrease of the ratio $r_0/a$ and the increase of $P_c/a^{3/2}$ with the increase of the indenter radius, consistently with experiments. The FFM coupled criterion has also been applied in \citep{Hahn2021} where the authors well reproduced the experimental results in \citep{Mouginot1985} using, however, different $G_c$ and $\sigmac$ compared to the values available in the literature.

Cone-shaped fracture has also been recently studied using the Phase-Field (PF) approach. This method relies on the seminal work by \citet{Francfort1998}, see also the subsequent developments in \citep{Bourdin2000, Amor2009, Miehe2010}. The PF method is based on Griffith's idea of the competition between the elastic and fracture energy counterparts: a pre-existing crack propagates if the increase in the surface energy necessary to create a new crack front is balanced by the reduction of the elastic strain energy stored in the body  \citep{Griffith1921}. In the PF variational approach, the fracture phenomenon is treated as nonlocal damage, and it is solved through an energy minimization which, in the $\Gamma$-convergence limit, consistently reproduces the Griffith theory of fracture. Therefore, a damage variable at the material point level is introduced as an additional primary unknown of the problem in addition to the displacement field. A recent review of the method can be found in \citep{Wu2020}.
The approach has been proved to successfully capture crack nucleation and propagation not only in isotropic linear elastic materials but also in the case of plasticity \citep{Ambati2015, MIEHE20161}, hyperelasticity \citep{Miehe2014, Russ2020, Mandal2020, Marulli2022} and heterogeneous composites \citep{Carollo2018, Guillen-Hernandez2020, Dean2020}. Experimental validation of two different variants of the PF method, with a rigorous convergence check of the staggered formulation, has been discussed in \citep{Cavuoto2022} in relation to PMMA specimen geometries with circular holes and V-notches, leading to complex patterns featuring crack nucleation, propagation, and even secondary propagation events.  

Concerning indentation problems, the PF approach has been applied to cone-shaped fracture in \citep{Strobl2019, Strobl2020, Wu2022, Kumar2022}, where the authors discussed the most appropriate phase-field formulation to simulate cracking of glass substrates due to flat-ended cylindrical punches, proposing a specific form of the tensile-compressive strain energy specific for the indentation tests \citep{Strobl2019, Strobl2020}. An alternative formulation has been given in \citep{Wu2022} with the so-called \textit{phase-field cohesive zone model} (PF-CZM), where the geometric crack function and the energetic degradation function have been adapted to reproduce a brittle fracture law with a linear softening curve. A revised PF approach can also be found in the work of \citet{Kumar2022}, where the authors introduced a strength of the glass boundary layer on the surface ($60 \megap$) different from that in the bulk ($150 \megap$), to account for microscopical defects located on the material surface, frequent in glass.

According to the above state of the art, it has to be noticed that less attention has been so far devoted to the case of spherical indentation, which has been discussed numerically using the PF approach only in \citep{Kindrachuk2020} for the case of a sphere with radius $1\millm$. The problem of spherical indentation requires, in fact, the solution of the contact problem in addition to the simulation of fracture in the bulk, which implies the solution of a strong nonlinear mechanical problem with two forms of nonlinearities. In \citep{Kindrachuk2020}, the continuous update of the contact area was addressed by solving the contact problem with a penalty method, coupling the mechanical and the phase-field evolution equations using a staggered solution scheme.

To make further progress on the prediction of cone cracks resulting from non-conforming contact, we propose here an innovative computational approach where we simulate the contact problem with the MPJR (eMbedded Profile for Joint Roughness) interface finite element formulation proposed in \citep{Reinoso2014}, and extended in \citep{Bonari2020, Bonari2022} for frictional and adhesive contact problems. The method allows considering the contact surfaces as nominally flat, but preserving the actual geometry of the indenter by embedding its geometry into an internal variable of the interface finite element to correct the normal gap function. As proved in the previous publications, the method can be successfully applied to 2D and 3D contact problems with a smooth shape of the indenter (spherical, wavy) and also to the more challenging contact problem involving a random fine-scale roughness. Therefore, this predictive methodology precludes the need to explicitly discretize the actual boundary geometry, which is, however, fully taken into account as an exact correction to the gap function.    

In this study, for the simulation of crack growth, the state-of-the-art version of the PF approach to fracture in \citep{Wu2022} is also implemented for the continuum substrate. \textcolor{black}{To make both formulations compatible, the nodal degrees of freedom of the MPJR interface finite elements are augmented to be consistent with the degrees of freedom of the finite elements used to discretize the continuum with the PF, which involve the nonlocal damage variable in addition to the displacement field components. This methodology also opens new perspectives in terms of modeling constitutive coupling effects between damage in the bulk surrounding the surface and the tribological properties, such as friction, wear, and adhesion. Moreover, the 3D formulation has also been detailed for 2D axisymmetric problems since they are particularly relevant to efficiently simulate spherical indentation fracture reducing the computation cost as compared to full-scale 3D simulations.}

The resulting overall computational formulation, equipped with both nonlinearities due to contact and fracture, is herein tested for the first time in relation to spherical indentation with smooth or rough spheres.
It is, in fact, known that another critical aspect often neglected in indentation-induced fracture simulations is the influence of surface morphology on the test results. In this concern, experimental results provide non-consistent trends depending on the indenter material.
In the case of steel indenters, the critical load and the ring crack radius in the abraded cases were greater than on the as-received surfaces, as reported in \citep{Conrad1979, Mouginot1985, Jyh-Woei1993, Lu1995}. Similarly, etching treatments in hydrofluoric acid of the glass surface increased the critical load $P_c$ with respect to the simply polished surface in \citep{HAMILTON1970}. 
Similarly, steel ball indentation on silicon wafer substrate in \citep{Jyh-Woei1993} showed that the critical load was higher for the polished silicon surface than for the abraded one, while the crack ring radius increased. 

On the other hand, in the case of a glass indenter on a glass substrate \citep{Johnson1973}, surface abrasion was reported not to affect the critical load.

Hence, we expect that mechanical properties and surface roughness resulting from different surface treatments might influence crack formations, and a computational framework accounting for surface roughness is desirable to gain insight into these issues.

Therefore, the present article aims at developing a new simulation framework with strong mechanical foundations that allows testing the potential of the PF approach coupled with the MPJR interface formulation for non-conformal indentation cracking problems. The article is structured as follows: Sec. \ref{sec:gov_eq} provides the overall coupled problem formulation, and it includes the treatment of the contact problem in Sec. \ref{sec:contact_form} and the phase-field formulation for fracture in Sec.\ref{sec:variational1}. \textcolor{black}{The axisymmetric MPJR method, presented for the first time in this work, is applied in a benchmark test to prove its efficiency against a standard node-to-segment approach in Section \ref{sec:benchmark}.} Section \tild\ref{sec:simulations} shows the application of the computational method to smooth and rough spherical indenters, \textcolor{black}{with novel insights into the physical problem}.
\FloatBarrier

\section {Governing equations of the coupled contact and fracture problem} \label{sec:gov_eq}

\begin{figure}[ht]
    \centering
    \includegraphics[width=0.6\linewidth]{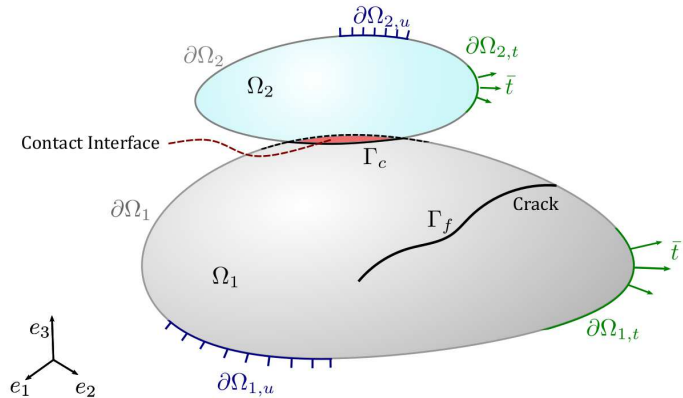}
    \caption{Non-conformal contact between two solids $\Omega_1$ and $\Omega_2$ that will lead to a contact domain $\Gamma_c$, function of the applied load or displacement, and an evolving crack $\Gamma_f \in \Omega_1$.}
    \label{fig:general}
\end{figure}

The present section describes the general variational framework of the system which includes: $(i)$ the contact interaction between two solid domains ($\Omega_i$ with $i=1,2$) at the contact interface $\Gamma_c$; and $(ii)$ the treatment of the brittle fracture problem. A crack $\Gamma_f$ is assumed to nucleate and propagate in the substrate, as shown in \fig\ref{fig:general}.

The free energy functional of the system reads:
\begin{equation}
    \Pi(\bu,\Gamma_f,\Gamma_c)=\Pi_{\Omega_1}+ \Pi_{\Omega_2} + \Pi_{\Gamma_f} + \Pi_{\Gamma_c}
\end{equation}
where $\Pi_{\Omega_1}$ and $\Pi_{\Omega_2}$ denote the total potential energy of the solids, $\Pi_{\Gamma_f}$ is the energy dissipated due to fracture, and $\Pi_{\Gamma_c}$ is the contribution due to the contact interactions.
Both contributions are discussed in the following subsections.

\subsection{Contact contribution to the weak form}\label{sec:contact_form}

This section focuses on the treatment of the contact problem between the indenter and the substrate. The MPJR interface finite element formulation introduced in \citep{Paggi2018} is herein adopted to discretize the interface. The formulation is herein specialized to frictionless and adhesiveless indentation problems, although the method also applies to such scenarios as shown in \citep{Bonari2022}. 

\begin{figure}[hp]
    \centering
    \includegraphics[width=0.6\linewidth]{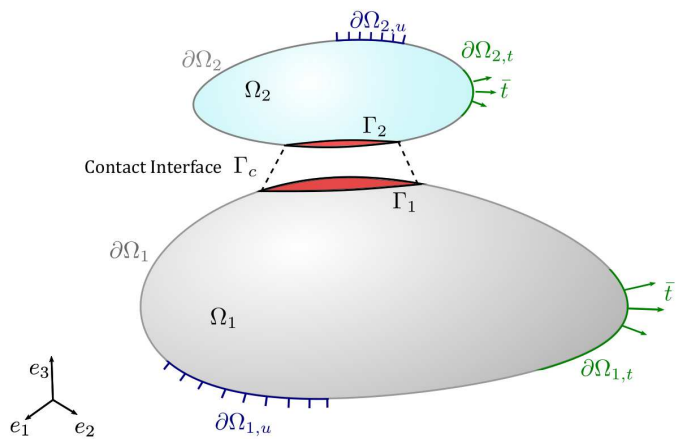}
    \caption{Solid domains $\Omega_1$ and $\Omega_2$ interacting through the contact interface $\Gamma_c$. }
    \label{fig:contact_solids}
\end{figure}

The proposed numerical method consists of a zero-thickness interface finite element separating the indenter and the substrate. Both materials can be modeled with their actual elastic properties, as in \citep{Reinoso2014}. It can also be applied to the case of a rigid indenter in contact with an elastic substrate as a special case. Let the two solids occupy the 3D domains $\Omega_i \in \mathbb{R}^3$ with $(i=1,2)$ in the undeformed configuration defined by the reference system $Oe_1e_2e_3$, as shown in \fig\ref{fig:contact_solids}. The position of a point in the body is given by the vector of its Cartesian coordinates $\bx$. The bodies are separated by an interface $\Gamma_c \in \mathbb{R}^2$ defined by the two opposite sides $\Gamma_1$ and $\Gamma_2$. 

Kinematic and traction boundaries conditions can be prescribed on disjointed parts of the solids' boundaries such that each boundary can be split into three parts: $(i)$ a portion where displacements are imposed, $\partial\Omega_{iu}$; $(ii)$ a portion where tractions $\hat{\mathbf{t}}$ are specified, $\partial\Omega_{it}$; $(iii)$ the contact interface $\Gamma_i$ where contact tractions are exchanged. Let the body $\Omega_1$ be the substrate with a smooth contact interface, while $\Omega_2$ is the indenter whose contact interface $\Gamma_2$ has an arbitrary shape that can be described by an analytical function (e.g., a parabolic or harmonic profile) or by a set of discrete data related to a more complex rough topology. 

\begin{figure}[h]
    \centering
    \includegraphics[width=0.5\linewidth]{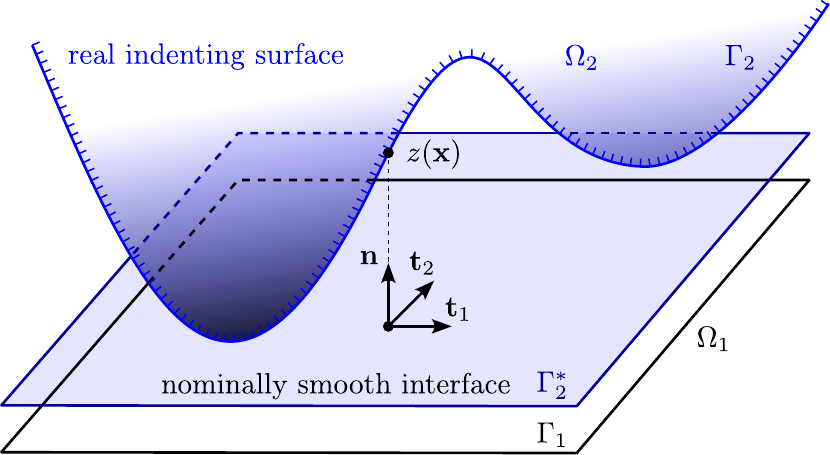}
    \caption{From original boundary to embedded  geometry}
    \label{fig:indenter_smooth}
\end{figure}

The core of the approach consists in simplifying the original boundary of the indenter $\Gamma_2$ into a smooth surface $\Gamma_2^*$, while the actual profile of the boundary is embedded point-wise in its exact form into the interface finite element formulation. The geometrical difference between $\Gamma_2$ and $\Gamma_2^*$ is mathematically described by a function $z(\bx)$ where $\bx=(x,y)^\text{T}$ is the coordinate position vector on the surface, see \fig\ref{fig:indenter_smooth}. 

This method reduces the contact problem to a nominally flat-to-flat interface $\Gamma_c^*$. It is solved by introducing the displacement field of the solids $\bu^{(i)} = (u^{(i)}, v^{(i)}, w^{(i)})^\text{T}$ such that the configuration of the system at the contact interface is described by the gap field $\bg=(g_n,g_{t1},g_{t2})^\text{T}$ defined as the projection of the relative displacement $\bu^{(1)} - \bu^{(2)}$ onto the normal and tangential directions of the interface:
\begin{equation}\label{eq2}
    g_n= \bn \cdot (\bu^{(1)} - \bu^{(2)}), \qquad
    g_{t1}= \bt_1 \cdot (\bu^{(1)} - \bu^{(2)}), \qquad
    g_{t2}= \bt_2 \cdot (\bu^{(1)} - \bu^{(2)})    
\end{equation}

The normal gap $g_n^*$ is then corrected to restore the exact shape of the indenter as follows:
\begin{equation}\label{eq:contact_correction}
    g_n^*=g_n+z(\mathbf{x})
\end{equation}

In the hypothesis of a frictionless and adhesiveless contact problem, the contact traction $\bp=(p_n, \tau_1,\tau_2)^\text{T}$ reduces to the normal component $p_n$. The associated contact conditions expressed using the standard Hertz-Signorini-Moreau inequalities read:
\begin{equation}
    g_n^*\ge 0, \qquad p_n\le 0, \qquad g_n^* p_n =0 \quad \text{on}~\Gamma_i\; 
\end{equation}
and are treated with a penalty approach \citep{Wriggers}[p. 118]: 
\begin{equation}\label{eq:contact}
    p_n= \begin{cases}
        0 \; &\text{if} \;  g_n^*\ge 0 \\
        k_p g_n^* \; &\text{if} \;  g_n^*<0
        \end{cases}
\end{equation}
where $k_p$ is the penalty stiffness.

Equation \eqref{eq:contact} gives a nonzero contact pressure for all the points within the nominally flat-to-flat interface where the corrected normal gap is negative valued. Since the exact geometrical corrective term $z(\mathbf{x})$ is provided at each integration point of the interface finite elements, the surface geometry is embedded directly inside the derivation of the system's stiffness matrix without the need for an explicit discretization of its geometry, which leads to a significant advantage in terms of modeling and discretization of complex surface topologies.

\textcolor{black}{In the case of spherical indentation problems, it is convenient to specialize the above 3D formulation by introducing the 2D axisymmetric model to devise a computationally effective method. In such a case, the displacement field vector of the $i$-th solid is reduced to two terms, $\mathbf{u}^{(i)}=(u_r^{(i)},v^{(i)})$, where $u_r$ and $v$ denote, respectively, the displacement in the radial direction and on the vertical direction. The gap vector also reduces to two terms, $\mathbf{g}=(g_n,g_{t})$, computed according to the first two equations \eqref{eq2}. The corrective term of the normal gap vector is still evaluated according to Eq.\eqref{eq:contact_correction}, where the position vector $\bx$ coincides with the radial coordinate $r$, such that $z(\bx)=z(r)$. The contact constraint conditions in Eq.\eqref{eq:contact} still apply, provided that the contact traction vector $\mathbf{p}=(p_n,\tau)^\text{T}$ with only two components is considered.}

\begin{figure}
    \centering
    \includegraphics[width=0.8\linewidth]{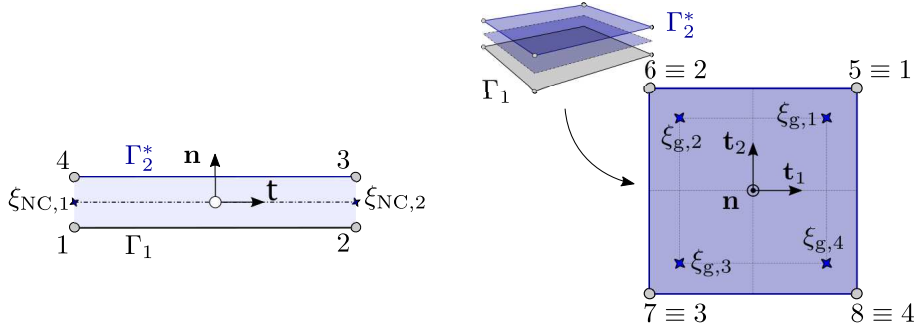}
    \caption{2D and 3D interface finite elements.}
    \label{fig:interface_elmts}
\end{figure}

The numerical treatment of the contact problem requires the discretization of the interface, \textcolor{black}{which is conducted with 8-nodes MPJR interface finite elements in 3D and with 4-nodes interface finite elements in 2D, see \fig\ref{fig:interface_elmts}}. 

A conformal mesh discretization is adopted for the continuum at the interface between the two solids. The solution method has, therefore, the same features as a segment-to-segment contact algorithm with fixed pairings \citep{PW}. 

Given the nodal displacement vector $\bar{\bu}=(u_1, v_1, w_1, ..., u_8, v_8, w_8)^\text{T}$ for a 3D problem, or $\bar{\bu}=(u_{r1}, v_1, ..., u_{r4}, v_4)^\text{T}$ for the 2D axisymmetric one, the normal gap inside the interface finite element can be derived by computing the relative displacement $\Delta \bu$ between the couples of nodes from the opposite sides of the interface using a matrix operator $\bL$. 
The result is then interpolated at any point inside the interface element through standard linear shape functions collected in the matrix operator $\bN$. Finally, the normal and tangential components of the gap are determined using a rotation matrix $\bR$ defined by the unit vectors $\bn$, $\bt_1$ and $\bt_2$ in 3D, or just $\bn$ and $\bt_1$ in 2D, both related to the local reference system of the interface finite element. In formulae, the discretized gap field can be written as:
\begin{equation}
    \bg=\bR \bN \bL \bu
\end{equation}

The deviation from planarity of the shape of the indenter profile can be taken into account by computing the corrected gap vector $\bg^*$ at each interface integration point, according to \equationame\eqref{eq:contact_correction}. 

If an analytical function is used to define the profile shape, then the correction $z(\bx)$ is computed by introducing the coordinates of the interface finite element nodes. Otherwise, if the surface/profile data are provided as a discrete set of elevations, as from data acquired from a profilometer or AFM, then those data are provided in input to the software. Such input data are stored in a history variable inside the user element routine only once, at the initialization of the problem. A mapping routine connects the external data to the proper node. The finite element discretization at the interface will be related to the spatial spacing of the external data. \textcolor{black}{If the phase-field problem requires a discretization of the continuum finer than the sampling spacing of the surface data field at the interface, which can be 
due to the internal length-scale constraint of the method (see Sec. 
 \ref{sec:variational1}), then a linear interpolation of the input heights field is performed to locally refine the data assigned to the nodes of the conformal interface discretization. Such an issue is not at stake in case of pure contact problems without fracture.}
\begin{figure}
    \centering
    \includegraphics[width=0.9\linewidth]{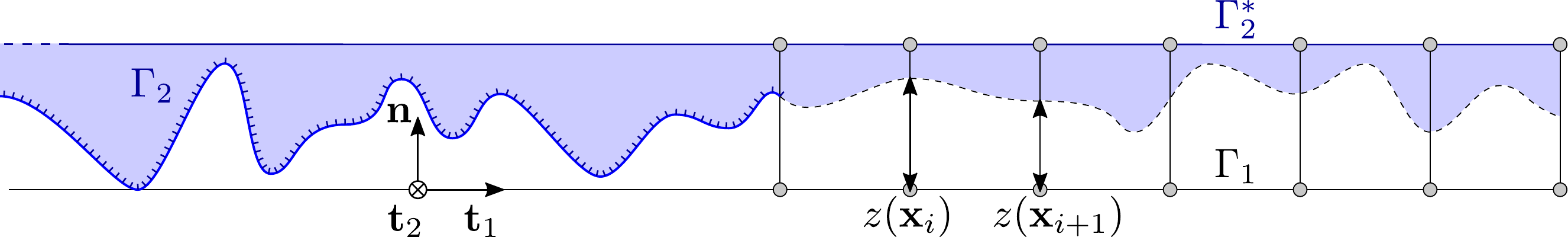}
    \caption{Element-wise profile discretization.}
    \label{fig:int_discr}
\end{figure}

The contribution of a single interface finite element to the variational formulation of the system is given by:
\begin{equation}\label{eq:contact_var}
    \delta \Pi = \int_{\Gamma_c^*} \delta g_n^* p_n  \dA,
\end{equation}
\textcolor{black}{where $\dA$ denotes the area of the surface patch in 3D, while it is equal to $2\pi r \dr$ for the 2D axisymmetric case.}

The equation leads to the following expressions for the element residual vector $\mathcal{R}_c$ and the element contact stiffness matrix $\mathcal{K}_c$ \textcolor{black}{associated with the mechanical field} to be used in a Newton-Raphson incremental-iterative solution scheme:
\begin{equation}\label{eq:resid_contact}
    \mathcal{R}_c=\int_{\Gamma_c^*} \bL^\text{T} \bN^\text{T} \bR^\text{T} \bt \dA, \qquad
    \mathcal{K}_c=\int_{\Gamma_c^*} \bL^\text{T} \bN^\text{T} \bR^\text{T} \mathbb{C} \bR \bN \bL \dA
\end{equation}
where $\mathbb{C}$ is the linearized interface constitutive matrix having only one nonzero component $\mathbb{C}_{11}=\partial p_n/\partial g_n=k_p$ for the contact points.

The combined phase-field interface finite element framework has been herein implemented using a monolithic fully implicit solution strategy in the finite element software FEAP8.6 \citep{Zienkiewicz2013}, following the methodology discussed in \citep{Linder}. \textcolor{black}{To make compatible the interface finite element with the phase-field finite element for the bulk which presents an additional degree of freedom associated to the damage variable in each node, the element residual vector and the element stiffness matrix contributions are expanded by mapping the corresponding terms in reference to the augmented vector of degrees of freedom $\bar{\bu^e}=(u_1, v_1, w_1, \phi_1, ..., u_8, v_8, w_8, \phi_8)^\text{T}$ for a 3D problem, or $\bar{\bu^e}=(u_{r1}, v_1, \phi_1, ..., u_{r4}, v_4, \phi_4)^\text{T}$ for the 2D axisymmetric case. Operatively, this is done by using a matrix operator $\mathbf{P}$ whose expression is collected in the Appendix for the 3D and the 2D cases, $\mathbf{P^{2D}}$ and $\mathbf{P^{3D}}$:
\begin{subequations}
\begin{align}
\mathcal{R}_c^e &=\mathbf{P}\mathcal{R}_c,\\
\mathcal{K}_c^e &=\mathbf{P}\mathcal{K}_c\mathbf{P}^\text{T}.
\end{align}
\end{subequations}
}

\subsection{phase-field contribution to the weak form}\label{sec:variational1}

The present subsection deals with the governing equations of the continuum solids in contact, which are treated according to the phase-field approach for fracture. The spherical indentation problem does not lead to micro-cracks, phase transformation in the materials, or plasticity, in contrast to sharp indentation. Hence, the tested materials will be considered linear elastic with nonlocal damage that, in the phase-field approach, tends to Griffith fracture in the limit of a vanishing internal length scale governing the nonlocality of the method. 

\begin{figure}[h!]
    \centering
    \subfloat[]{\includegraphics[width=0.45\textwidth]{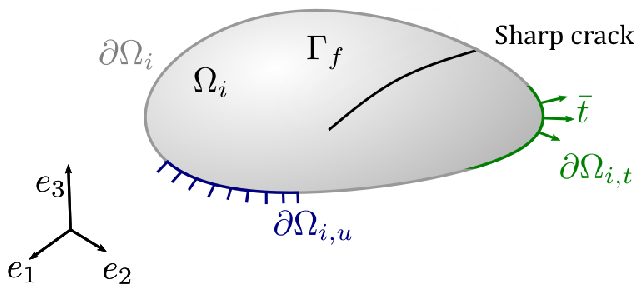}}\qquad
    \subfloat[]{\includegraphics[width=0.45\textwidth]{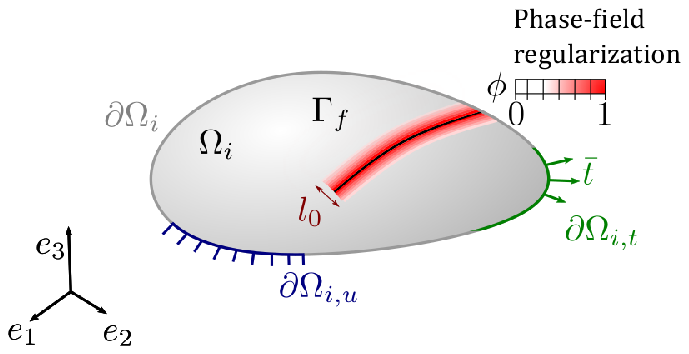}}
    \caption{A cracking solid $\Omega_i$ with a sharp crack $\Gamma_f$ in (a) and its phase-field nonlocal regularization in (b).}
    \label{fig:solid}
\end{figure}

The variational formulation is developed considering a standard solid domain $\Omega_i\subset \mathbb{R}^3$ with $i=1,2$ having boundaries denoted by $\partial\Omega_i$, see \fig\ref{fig:solid}.

It is also assumed that the arbitrary solid is subjected to specific body forces $\bar{\mathbf{b}}$, imposed displacements $\bar{\bu}$ on $\Omega_{i,t}$, and boundary tractions $\bar{\mathbf{t}}$ on $\Omega_{i,t}$, possibly caused by contact.
The external potential energy of the system reads:
\begin{equation}
    \mathcal{P}=\int_{\Omega_i\setminus\Gamma_f} \bar{\mathbf{b}} \cdot \bu \dV + \int_{\partial \Omega_{i,t}} \bar{\mathbf{t}} \cdot \bu \dA
\end{equation}
and the free energy functional of the systems is: 
\begin{equation}\label{eq:strain_energy_tot}
  \Pi(\bu,\Gamma_f)=\int_{\Omega_i\setminus\Gamma_f} \psi(\bepsilon) \dV + \int_{\Gamma_f} \Gc \dA
\end{equation}
Where $\psi(\bepsilon)$ is the elastic energy density of the body written in terms of the strain field $\bepsilon$. The integral $\int_{\Gamma_f} \Gc \dA$ identifies the energy dissipation due to fracture events at the crack set $\Gamma_f$, while $\Gc$ is the critical energy release rate or fracture toughness of the bulk material. This integral cannot be directly evaluated because the crack set $\Gamma_f$ is apriori unknown. According to the variational approach to fracture proposed in \citet{Francfort1998}, the problem is solved by substituting the sharp crack with a transition region from undamaged to broken material. The sharp crack is therefore approximated as a band of finite width $l_0$ characterized by a crack phase-field parameter ${\phi \in \left[0,1 \right]} $ such that $\phi=0$ denotes the intact material and  $\phi=1$ represents the cracked one. The crack approximation converges to the sharp crack when such a band thickness approaches zero.

The energy contribution due to fracture is obtained through its smeared nonlocal approximation:
 \begin{equation}
   \int_{\Gamma_f}\Gc \dA \approx \int_{\Omega_i}\Gc \gamma(\phi;\nabla \phi) \dV  
 \end{equation}
 
Accordingly, the free energy functional expressed in \equationame\tild\eqref{eq:strain_energy_tot} becomes
\begin{equation}
\label{eq:en_fun_pf}
\Pi(\bu;\phi)=\int_{\Omega_i} g(\phi)\psi(\bepsilon)  \dV + \int_{\Omega_i}\Gc \gamma(\phi;\nabla \phi) \dV
\end{equation}
where $g(\phi)$ is called \emph{energetic degradation function} and acts to reduce the elastic stiffness of the material.
  
There are different choices for the function $g(\phi)$. In this work, we used the model introduced by Bourdin et al. in \citet{Bourdin2000}: 
\begin{equation}
    g(\phi)=(1-\phi)^2+k_{\text{res}}    
\end{equation}
Where a small positive parameter $k_{\text{res}}$ is used to avoid numerical instabilities at the fully cracked state. In \equationame\tild\eqref{eq:en_fun_pf}, $\gamma(\phi;\nabla \phi)$ is the crack surface density function that assumes this generic form for the AT2 approach \citep{Miehe2010}:
 \begin{equation}
     \gamma(\phi,\nabla \phi)=\frac{1}{2} \left( \frac{\phi^2}{l_0}+l_0 |\nabla \phi|^2 \right)     
 \end{equation}
where $l_0$ is the length scale that defines the width of the diffusive crack band as shown in \fig\ref{fig:solid}b.

The reader is referred to \citet{Amor2009}, \citet{Tanne2018}, \citet{Strobl2020}, and \citet{Cavuoto2022} for a specific discussion on how to identify the value of $l_0$ from experimental data.  

According to the above state-of-the-art literature, the following variational formulation with respect to the primary field $\bu$ and $\phi$ can be derived:
\begin{subequations}
\begin{align}
     \delta \Pi_u = & \int_{\Omega_i} g(\phi) \bsigma(\bu) : \bepsilon(\delta \bu) \dV\label{eq:variationala} 
     \\
     \delta\Pi_\phi = & \int_{\Omega_i} \dfrac{\text{d} g(\phi)}{\text{d} \phi} \psi(\bepsilon)\delta \phi \dV -
     \int_{\Omega} \Gc \left\{ \frac{\phi}{l_0} \delta \phi + l_0 \nabla \phi \cdot \nabla  \delta \phi \right\} \dV\label{eq:variationalb}
\end{align}\label{eq:variational}
\end{subequations} 
where $\delta \bu$ and $\delta \phi$ stand for the virtual variation of the displacements and the phase-field, $\bsigma(\bu)$ is the Cauchy stress tensor for the undamaged configuration defined as $\bsigma=\mathbb{C}_0:\bepsilon$ with $\mathbb{C}_0$ the standard fourth-order stiffness tensor of an isotropic linear elastic material.

\textcolor{black}{The strain and stress tensors in Voigt notation for the 3D and the 2D axisymmetric models read, respectively:}
\begin{equation}
    \textcolor{black}{\bepsilon=\left[ \epsilon_{x}, \epsilon_{y}, \epsilon_{z}, \gamma_{xy}, \gamma_{yz}, \gamma_{xz} \right]^\text{T}, \qquad \bepsilon=\left[ \epsilon_r, \epsilon_z, \epsilon_\theta, \gamma_{rz} \right]^\text{T}}
\end{equation}

\begin{equation}
    \textcolor{black}{\bsigma=\left[ \sigma_{x}, \sigma_{y}, \sigma_{z}, \tau_{xy}, \tau_{yz}, \tau_{xz} \right]^\text{T}, \qquad \bsigma=\left[ \sigma_r, \sigma_z, \sigma_\theta, \tau_{rz} \right]^\text{T}}
\end{equation}

In \equationame\tild\eqref{eq:variationalb}, the elastic strain energy has to be split into its positive and negative counterparts to avoid crack propagation in compression: $\psi(\bepsilon)=\psi_{+}(\bepsilon)+\psi_{-}(\bepsilon)$. Considering only the positive contribution, the same equation becomes:
\begin{equation} \label{eq:var_plus1}
    \delta\Pi_\phi = \int_{\Omega_i} \dfrac{\text{d} g(\phi)}{\text{d} \phi} \psi_{+}(\bepsilon)\delta \phi \dV -
     \int_{\Omega} \Gc \left\{ \frac{\phi}{l_0} \delta \phi + l_0 \nabla \phi \cdot \nabla  \delta \phi \right\} \dV
\end{equation}

The energy driving the crack growth $\psi_{+}$ has been derived as in \citet{Strobl2020}:
\begin{equation}
    \psi_{+}:=\frac{1+\nu}{2 E} \left( \langle \sigma_1 \rangle_{+} ^2 + \langle \sigma_2 \rangle_{+} ^2 + \langle \sigma_3 \rangle_{+} ^2 \right) - \frac{\nu}{2E} \langle \text{tr} (\bsigma) \rangle_{+} ^2
\end{equation}
where $\langle  \square \rangle$ is the Macaulay bracket operator $\langle  \square \rangle_{+}:= (|\square|+\square)/2$, and $\sigma_i$ are the eigenvalues of the stress tensor for the intact material. 

The irreversibility of the phase-field evolution has been enforced by following the approach in \citet{Miehe2010} by using a history-field variable of the maximum positive energy contribution:
\begin{equation}
    \mathcal{H}(\bu,t)= \max_{t \in [0,T]} \psi_{+}(\bepsilon)
\end{equation}
where $t$ is the current pseudo-time step in a quasi-static simulation. With this assumption, \equationame\eqref{eq:var_plus1} becomes:
\begin{equation} \label{eq:var_plus}
    \delta\Pi_\phi = \int_{\Omega_i} \dfrac{\text{d} g(\phi)}{\text{d} \phi} \mathcal{H} \delta \phi  \dV -
     \int_{\Omega} \Gc \left\{ \frac{\phi}{l_0} \delta \phi + l_0 \nabla \phi \cdot \nabla  \delta \phi \right\} \dV
\end{equation}

Isoparametric finite elements with standard bilinear shape functions $N^I(\mathbf{\xi})$ have been used for the spatial discretization of the domain (see \citet{Msekh2015} for more details). The approximated displacement field, the phase-field, and their variations read:
\begin{equation}
    \bu=\sum_{I=1}^{nd} N^I \bu_I, \qquad 
    \delta\bu=\sum_{I=1}^{nd} N^I \delta\bu_I, \qquad
    \phi=\sum_{I=1}^{nd} N^I \phi_I, \qquad
    \delta\phi=\sum_{I=1}^{nd} N^I \delta \phi_I
\end{equation}
where $nd$ stands for the number of nodes for each finite element, $\bu_I$ and $\phi_I$ denote the nodal values of the displacement and phase-field, respectively, which are collected in the corresponding vectors $\bar{\bu}$ and $\bar{\phi}$.

The strain field is interpolated through the displacement-strain operator $\bB_u$, while the gradient of the phase-field via $\bB_{\phi}$:
\begin{equation}\label{eq:epsilon}
    \bepsilon=\bB_u \bu, \qquad
    \nabla_{\bx} \phi= \bB_\phi \phi 
\end{equation}
\textcolor{black}{where $\bB_u$ needs to be specialised for the 3D and the 2D axisymmetric cases:}
\begin{equation}
\textcolor{black}{\bB_u=\begin{bmatrix}
      \dfrac{\partial N^I}{\partial x} & 0 & 0\\
      0 & \dfrac{\partial N^I}{\partial y} & 0\\
      0 & 0 & \dfrac{\partial N^I}{\partial z}\\
    \dfrac{\partial N^I}{\partial y} & \dfrac{\partial N^I}{\partial x} & 0 \\
    0 & \dfrac{\partial N^I}{\partial y} & \dfrac{\partial N^I}{\partial z} \\
    \dfrac{\partial N^I}{\partial z} & 0 & \dfrac{\partial N^I}{\partial x} \\    
      \end{bmatrix}, \qquad
      \bB_u=\begin{bmatrix}
      \dfrac{\partial N^I}{\partial r} & 0 \\
      0 & \dfrac{\partial N^I}{\partial z} \\
    \dfrac{N^I}{r} & 0 \\
    \dfrac{\partial N^I}{\partial z} & \dfrac{\partial N^I}{\partial r}
      \end{bmatrix}}
\end{equation}
while $\bB_{\phi}$ for the 3D and the 2D axisymmetric cases read:
\begin{equation}
\textcolor{black}{\bB_\phi=\begin{bmatrix}
      \dfrac{\partial N^I}{\partial x}\\
      \dfrac{\partial N^I}{\partial y}\\
      \dfrac{\partial N^I}{\partial z}    
      \end{bmatrix}, \qquad
      \bB_\phi=\begin{bmatrix}
      \dfrac{\partial N^I}{\partial r} \\
      \dfrac{\partial N^I}{\partial z} \\
      \end{bmatrix}}
\end{equation}

With the previous interpolation schemes, the residual vectors $\mathcal{R}_u$ and $\mathcal{R}_\phi$ associated with the displacement and the phase-field, respectively read:
\begin{subequations}
\begin{align}
    \mathcal{R}_u    &=\int_{\Omega_i} \left[(1-\phi)^2+k_{res}\right]  
                        \bB_u^\text{T} \bsigma \dV \\
    \mathcal{R}_\phi &= \int_{\Omega_i} -2(1-\phi) \bN^T \mathcal{H} \dV + 
                        \int_{\Omega_i} \Gc l_0 \left[ \bB_\phi^\text{T} \nabla_x \phi + \frac{1}{l_0^2} \bN^\text{T} \phi \right] \dV
\end{align}
\end{subequations}
\textcolor{black}{where the volume integration $dV$ is equal to $2\pi r \dr \text{dz}$ for the axisymmetric setting.}
The expressions of the element stiffness matrices necessary for linearizing the resulting nonlinear system can be found in \citet{Msekh2015}.

\FloatBarrier
\section{\textcolor{black}{Benchmark tests: MPJR method for axisymmetric contact problems}} \label{sec:benchmark}

\textcolor{black}{The MPJR method described in \sectionname\ref{sec:contact_form}  embeds the spherical (in 3D) or circular (in 2D axisymmetric problems) indenter geometry into the interface elements, in order to retrieve the exact solution of the actual problem without the need of explicitly discretizing its shape. For a benchmark test, the reader is referred to \citep{Bonari2022} for the Hertzian contact problem between a cylinder and a plane, with friction, under plane strain conditions. }

\textcolor{black}{In the present work, the methodology has been extended to 2D axisymmetric contact problems with spherical indenters, with the aim of studying the problem of indentation-induced cracks in the substrate. The contact solution obtained with the MPJR method, for a radius of the sphere $R_s=10 \millm$, is compared with: $(i)$ the Hertzian analytical solution for a half-space; $(ii)$ the numerical solution obtained by explicitly discretizing the spherical indenter following exactly its shape and solving the contact problem with the standard node-to-segment contact algorithm, with the same value of the penalty parameter.}

\begin{figure}[h]
    \centering    \includegraphics[width=0.9\linewidth]{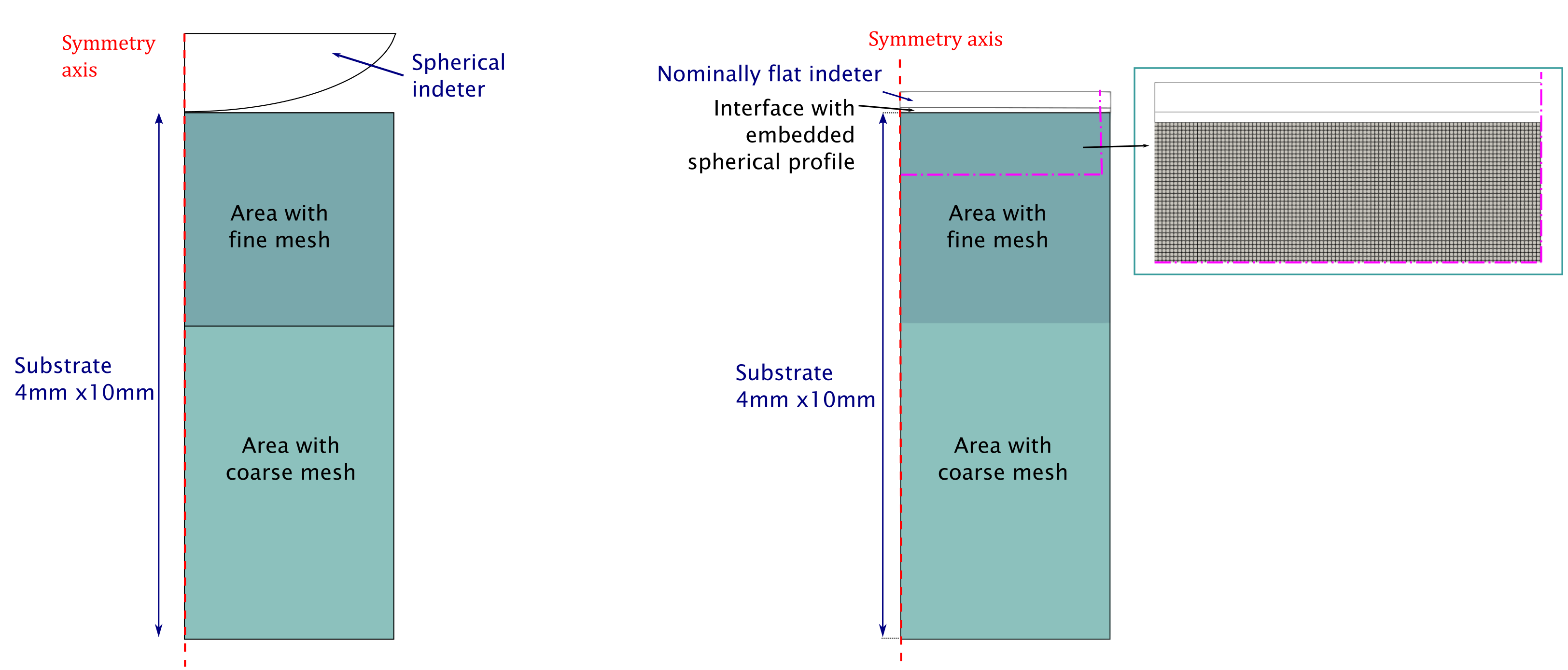}
    \caption{\textcolor{black}{Finite element model with the indenter explicitly discretized for the standard node-to-segment algorithm on the left and the model with the MPJR interface finite elements on the right. The same substrate mesh has been used in both models.} }
    \label{fig:model_comp}
\end{figure}

\textcolor{black}{The FEM models used for the standard spherical contact problem and the MPJR method are presented in \fig\ref{fig:model_comp}.}

\textcolor{black}{The simulation has been carried out under displacement control, with a maximum displacement perpendicular to the substrate of $0.01 \millm$, achieved in 100 steps. The substrate size ($4 \millm$ wide and $10 \millm$ deep) is large enough with respect to the contact radius according to Hertz's theory. Regarding the mechanical properties, the Young's modulus of the substrate is $63.40 \gigap$, and the Poisson ratio is $0.2$, while the indenter Young's modulus has been set 100 times higher than that of the substrate to simulate a rigid one. No damage has been considered for the solids in this benchmark test. The penalty parameter has been set equal to $1\times 10^9$ N/mm.}

\begin{figure}[h!]
    \centering
    \includegraphics[width=0.8\linewidth]{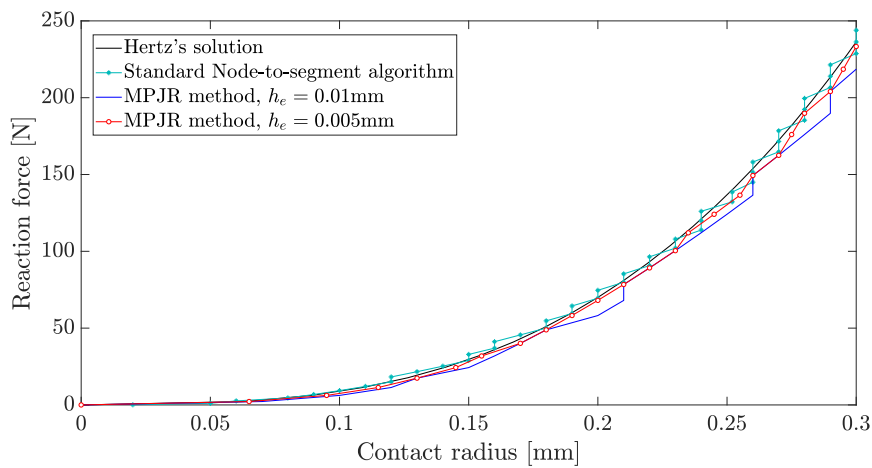}
    \caption{\textcolor{black}{Comparison between the Hertz equation (black line) and the result of the axisymmetric simulations obtained with a standard node-to-segment contact algorithm and the MPJR interface finite elements and two different mesh size, $h_e$.}}
    \label{fig:comparison}
\end{figure}

\begin{figure}[h!]
    \centering
    \includegraphics[width=0.8\linewidth]{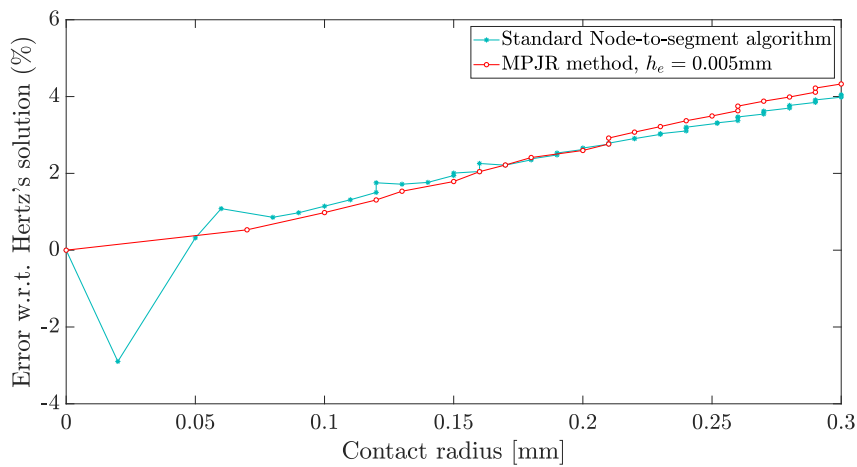}
    \caption{\textcolor{black}{Reaction force error with respect to Hertz solution obtained using a standard FEM axisymmetric model with a node-to-segment contact algorithm and the MPJR method with the same mesh size.}}
    \label{fig:error}
\end{figure}

\textcolor{black}{Results of the comparison shown in Figs. \ref{fig:comparison} and \ref{fig:error} highlight the excellent performance of the MPJR method, which also outperforms the standard node-to-segment contact algorithm, which presents some larger error for the first pseudo-time step, due to the contact search algorithm which is avoided in the MPJR formulation. The solution obtained with the MPJR method for two uniform FE discretizations and different mesh sizes, $h_e$, shows that the accuracy can be improved by mesh refinement, as shown in \fig\ref{fig:comparison} where the solutions for elements size $h_e=0.01\millm$ and $h_e=0.005\millm$ have been compared.}

\section{Simulation of Hertzian cone cracks for smooth or rough spheres} 
\label{sec:simulations}

\begin{figure}[h!]
    \centering
    \subfloat[Real geometry.]{
    \includegraphics[width=0.45\linewidth]{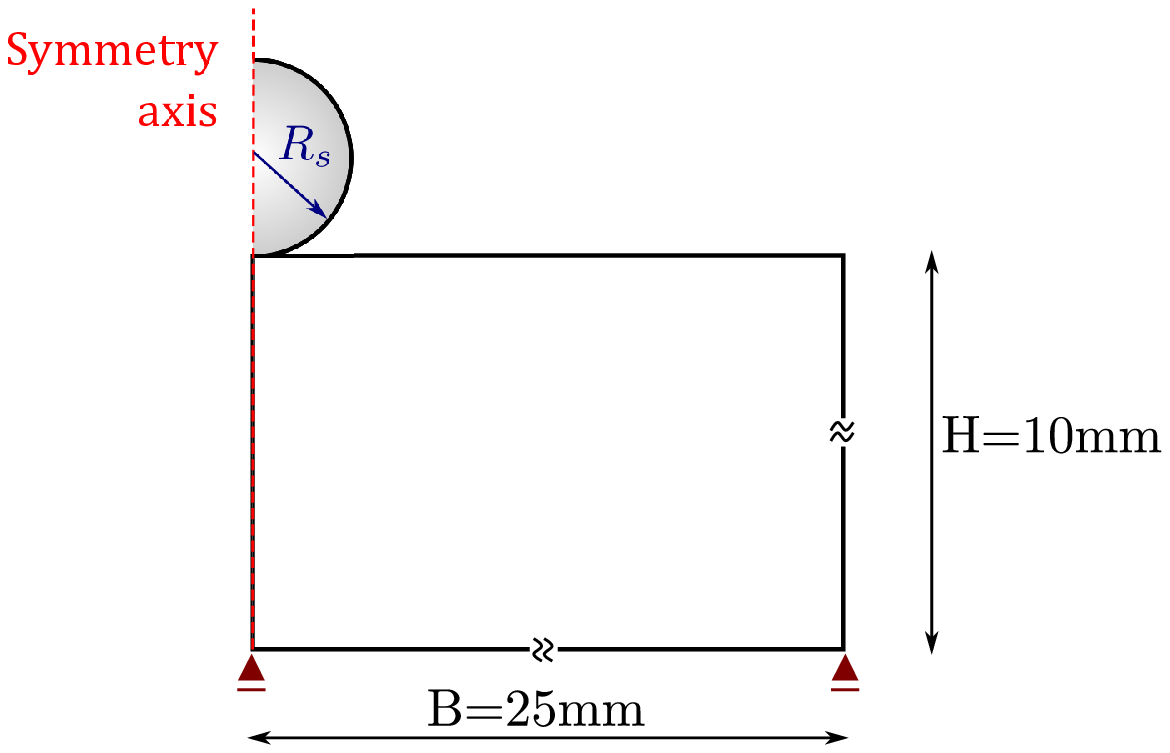}}
    \,
    \subfloat[Model geometry.]{
    \includegraphics[width=0.47\linewidth]{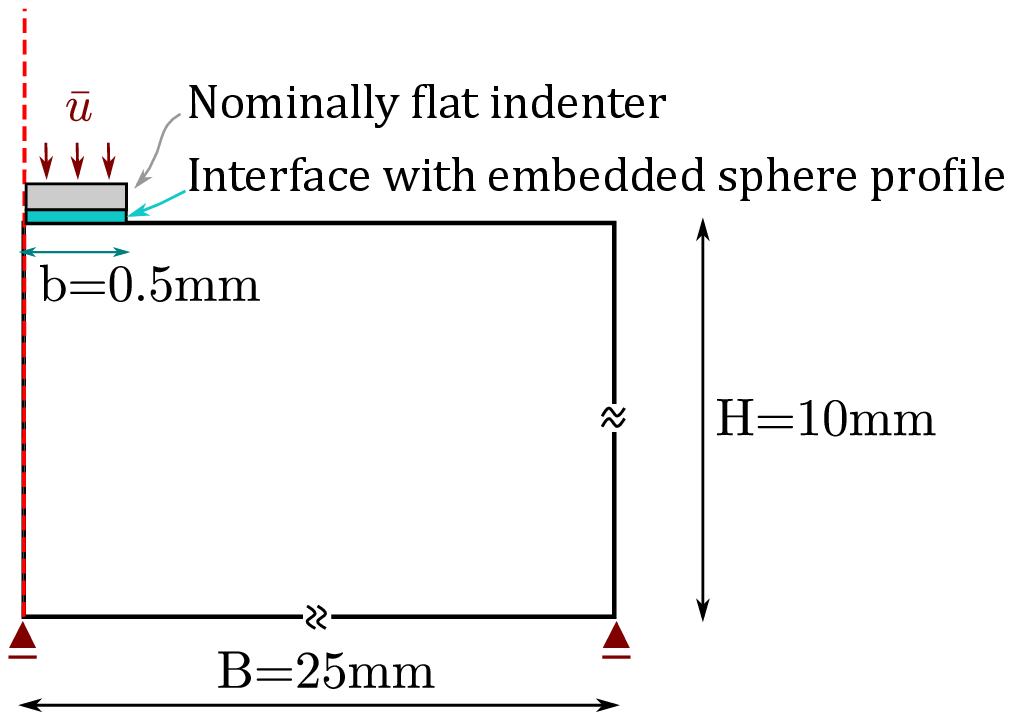}}
    \caption{\textcolor{black}{Real geometry (a) and equivalent model geometry based on the MPJR interface finite elements embedding the actual shape of the spherical indenter (b).}}
    \label{fig:geom}
\end{figure}

The Hertzian indentation test has been simulated by exploiting the axial symmetry of the model in a quasi-static framework. 
The model geometry and boundary conditions are sketched in \fig\ref{fig:geom}, which compares the real geometry of the test with the model geometry which employs the MPJR interface finite elements \textcolor{black}{embedding the actual spherical shape in the flat indenter}. The model of the substrate consists of a $25 \times 10 \millm$ rectangular domain as in the experimental studies in \citep{Conrad1979, Jyh-Woei1993}, which exploits symmetry conditions along the vertical axis on the left.

\begin{figure}[h!]
    \centering
    \includegraphics[width=0.9\linewidth]{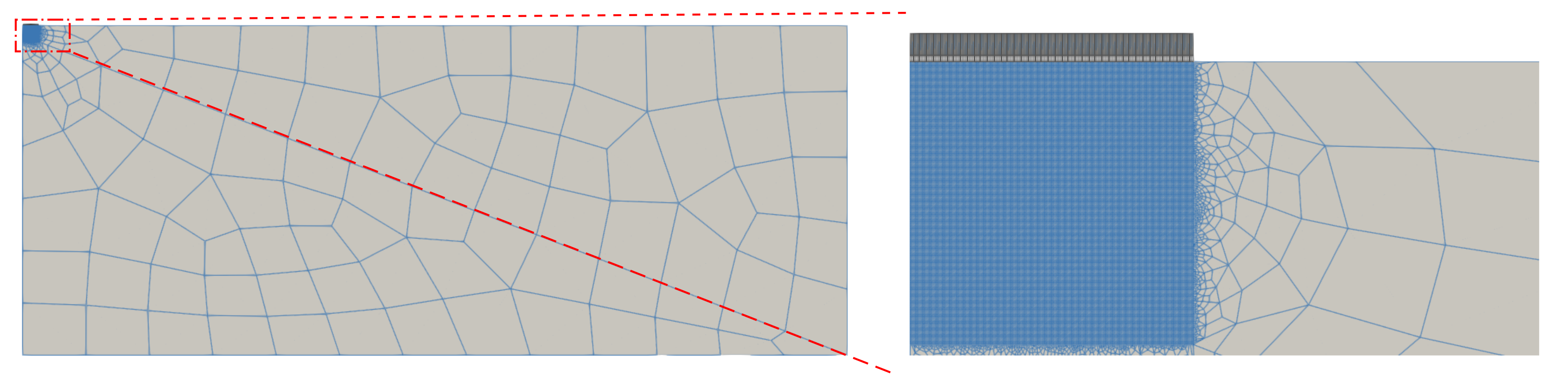}
    \caption{Discretization of the indentation test with the magnification of the refined area having a characteristic elements size $h_e=0.0025\millm$.}
    \label{fig:mesh}
\end{figure}

\textcolor{black}{Thanks to the MPJR methodology, the indenter has been modeled as a nominally flat punch using a layer of interface finite elements embedding the exact spherical profile} as an analytical function of the interface coordinates: $z(\mathbf{x})=R_s - \sqrt{R_s^2-\lVert \mathbf{x} \rVert}$, where $R_s=1\millm$.
The simulation is carried out under displacement control up to a value leading to a maximum contact radius $a$ smaller than the radius of the sphere. Therefore, for this model, the portion of the interface embedding the spherical profile has been set equal to $0.5\millm$, corresponding to the size of the refined FE discretization in \fig\ref{fig:mesh}, which is also much larger than the maximum contact radius reached in the simulation.

The material properties of the soda-lime glass substrate have been taken from \citep{Conrad1979} and reported in \tab\ref{tab:prop_jyh}. The fracture energy of the glass has been estimated to $0.009 \,\si{N/mm}$ in \citep{Mouginot1985}. 

\textcolor{black}{A discussion on the appropriate value for $l_0$ and the effect of the choice on the crack pattern can be found in \cite{Strobl2019}, \citep{Wu2022}, where different values of $l_0$ and glass tensile strength ranging between $50 \megap$ and $150 \megap$ have been tested in the simulations and compared with experimental data for the case of flat-ended cylindrical indenters. The authors showed that the critical displacement required for the crack nucleation increases with $\sigmac$, while the ring radius $r_0$ decreases when the tensile strength increase. }

\textcolor{black}{Moreover, according to \cite{Strobl2020}, the length scale parameter has to satisfy another criterion: it has to be chosen small enough to capture the extension of a spontaneous ring crack on the surface given in  ($l_0 \le 0.04 \millm$).}

\textcolor{black}{Considering both aspects, the length scale parameter has been set equal to $0.01 \,\millm$; it corresponds to a tensile strength of the glass of $77 \megap$ according to the well-known formula for the AT2 phase-field approach to correlate the strength to the other model parameters: $\sigma_c=\sqrt{\frac{27}{256} \frac{E G_c}{l_0}}$. As shown in the following paragraph, the value $l_0=0.01\millm$ allows a good reproduction of the experimental trends, and the phase-field discretization of the model has been refined where cracks are expected to nucleate (element size $h_e \le l_0/4=0.0025\millm$).}

\begin{table}[h]
    \centering
    \begin{tabular}{|c|c|c|c|c|c|}
         \hline
         &  Young's Modulus & Poisson Ratio & Fracture energy & Tensile strength & Length scale \\
         \hline
        Glass & $63.40 \gigap$ & $0.20$ & $0.009 \,\si{N/mm}$ & $77\megap$ & $0.01\millm$ \\
        \hline
    \end{tabular}
    \caption{Mechanical and fracture properties of the substrate taken from \cite{Conrad1979, Mouginot1985}.}
    \label{tab:prop_jyh}
\end{table}

\begin{figure}[h!]
    \centering
    \includegraphics[width=0.5\linewidth]{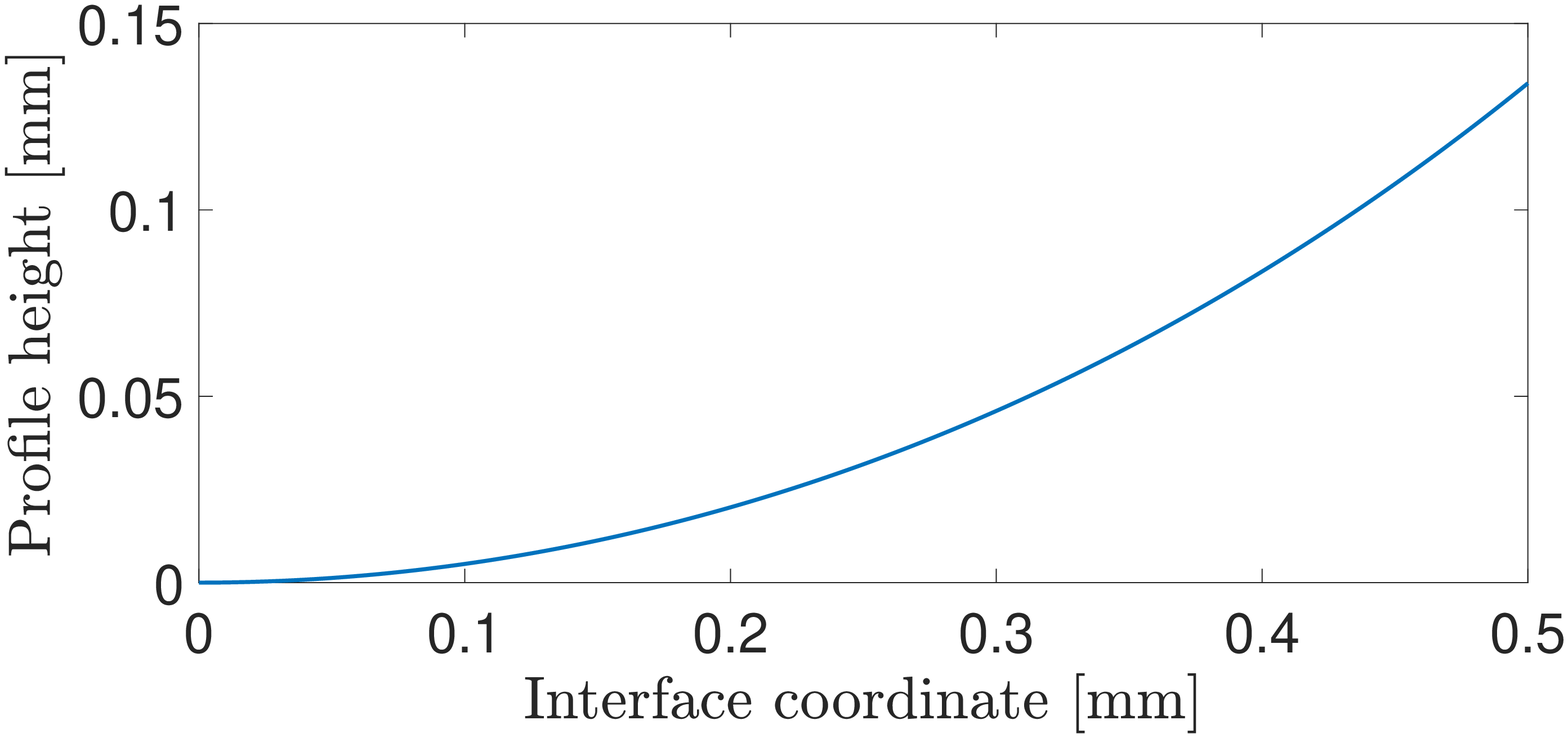}
    \caption{Profile of the spherical indenter with radius $1\millm$.}
    \label{fig:profile}
\end{figure}

\begin{figure}[h!]
    \centering
    \subfloat[$\bar{u}=5.2\times10^{-3}\millm$]{
    \includegraphics[width=0.4\linewidth]{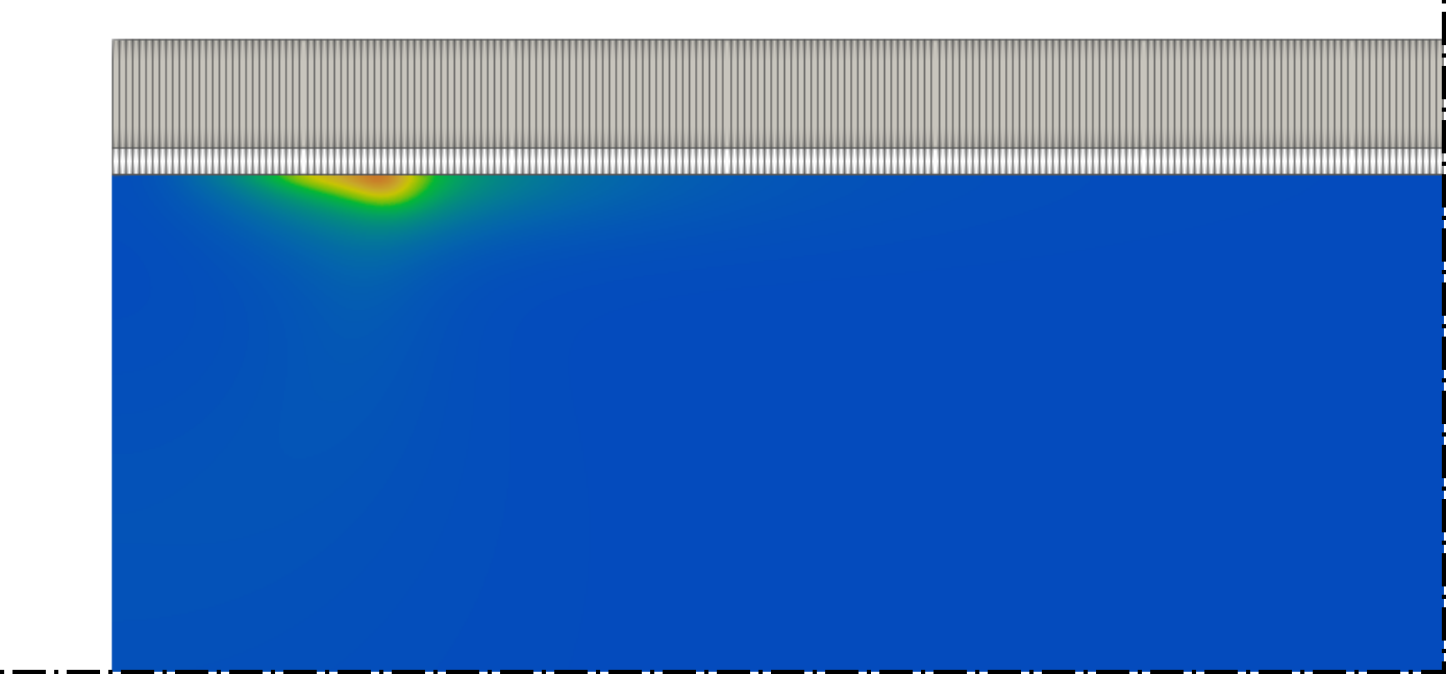}}
    \,
    \subfloat[$\bar{u}=5.4\times10^{-3}\millm$]{
    \includegraphics[width=0.4\linewidth]{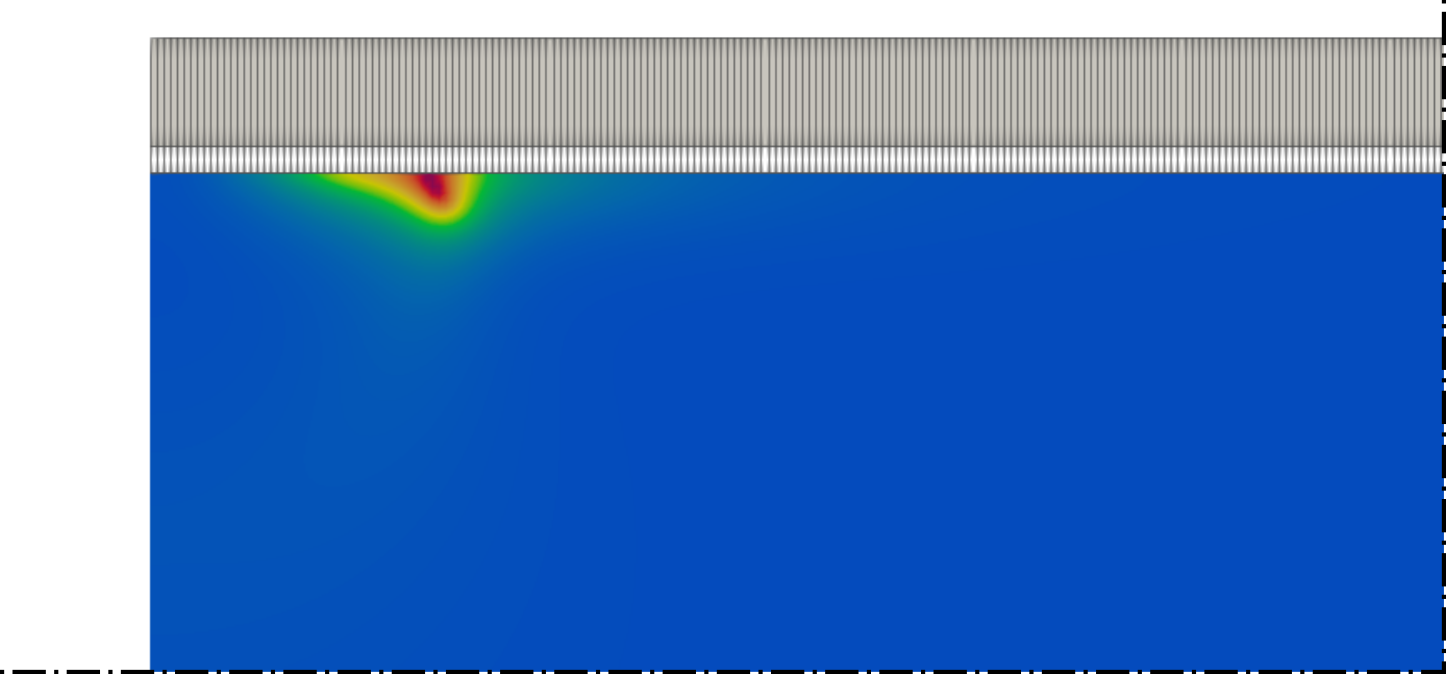}}
    \\
    \subfloat[$\bar{u}=6.5\times10^{-3}\millm$]{
    \includegraphics[width=0.4\linewidth]{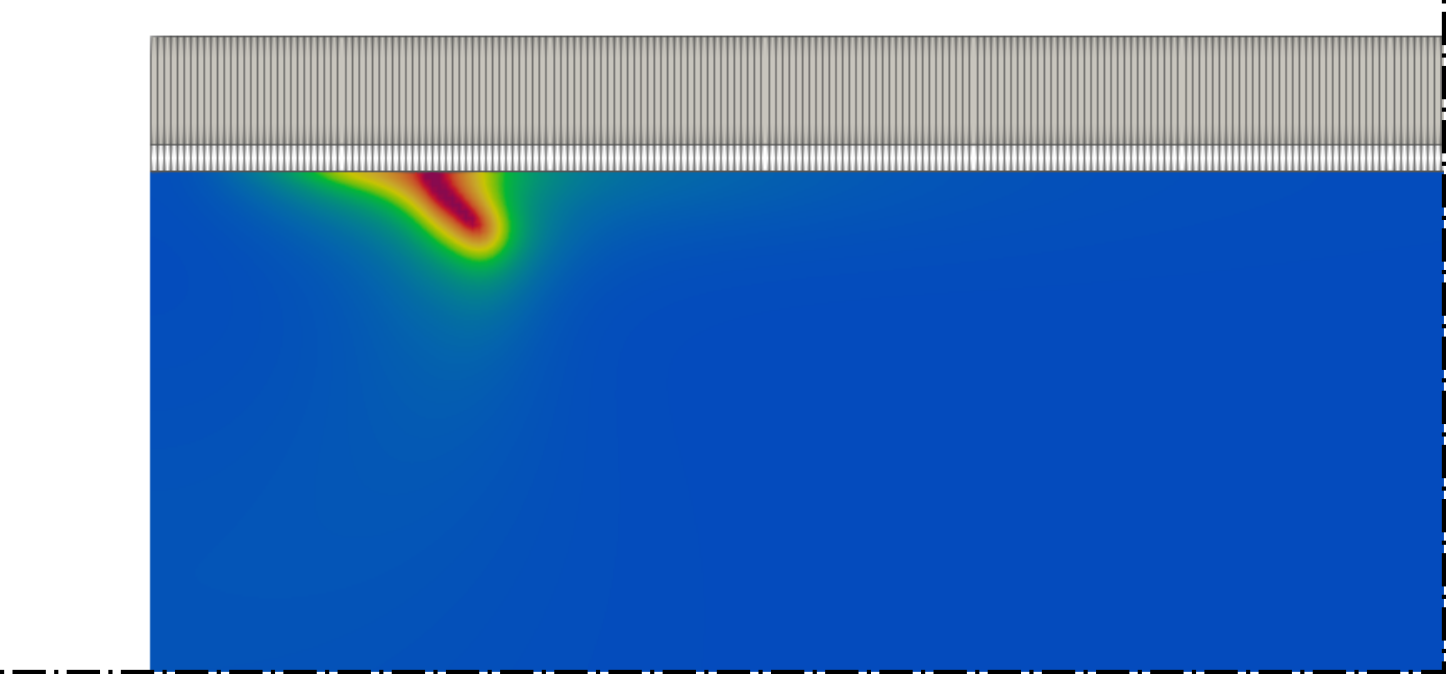}}
    \,    
    \subfloat[$\bar{u}=1.\times10^{-2}\millm$]{
    \includegraphics[width=0.4\linewidth]{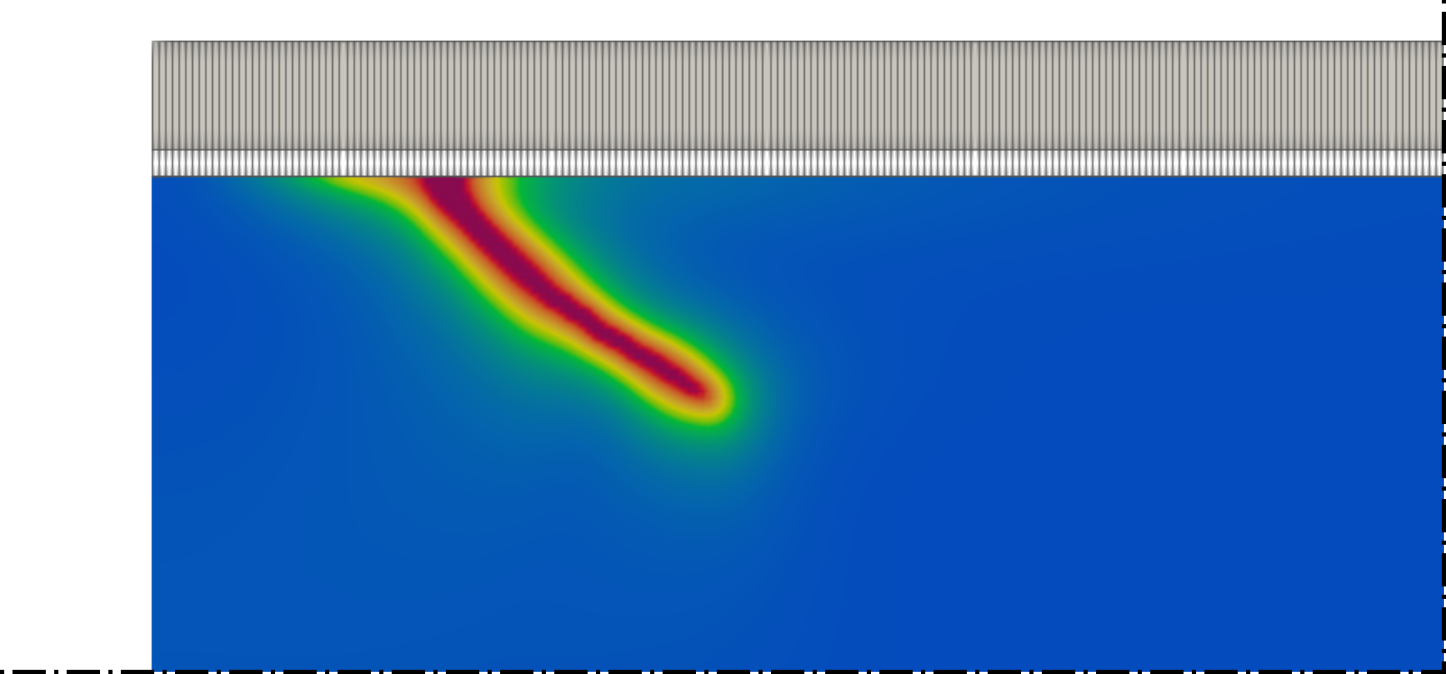}}\\
    \includegraphics[width=0.2\linewidth]{legend_horizontal.pdf}\\
    \caption{Snapshots of the fracture evolution for the indentation test with a sphere radius $1\millm$.}
    \label{fig:crack_evol}
\end{figure}

As already stated, we remark here that, even though the model in \fig\ref{fig:crack_evol} shows a flat indenter, the interface finite elements exactly embed the spherical profile shown in \fig\ref{fig:profile}. 

\begin{figure}[h!]
    \centering
    \subfloat[$\bar{u}=5.2\times10^{-3}\millm$]{
    \includegraphics[width=0.4\linewidth]{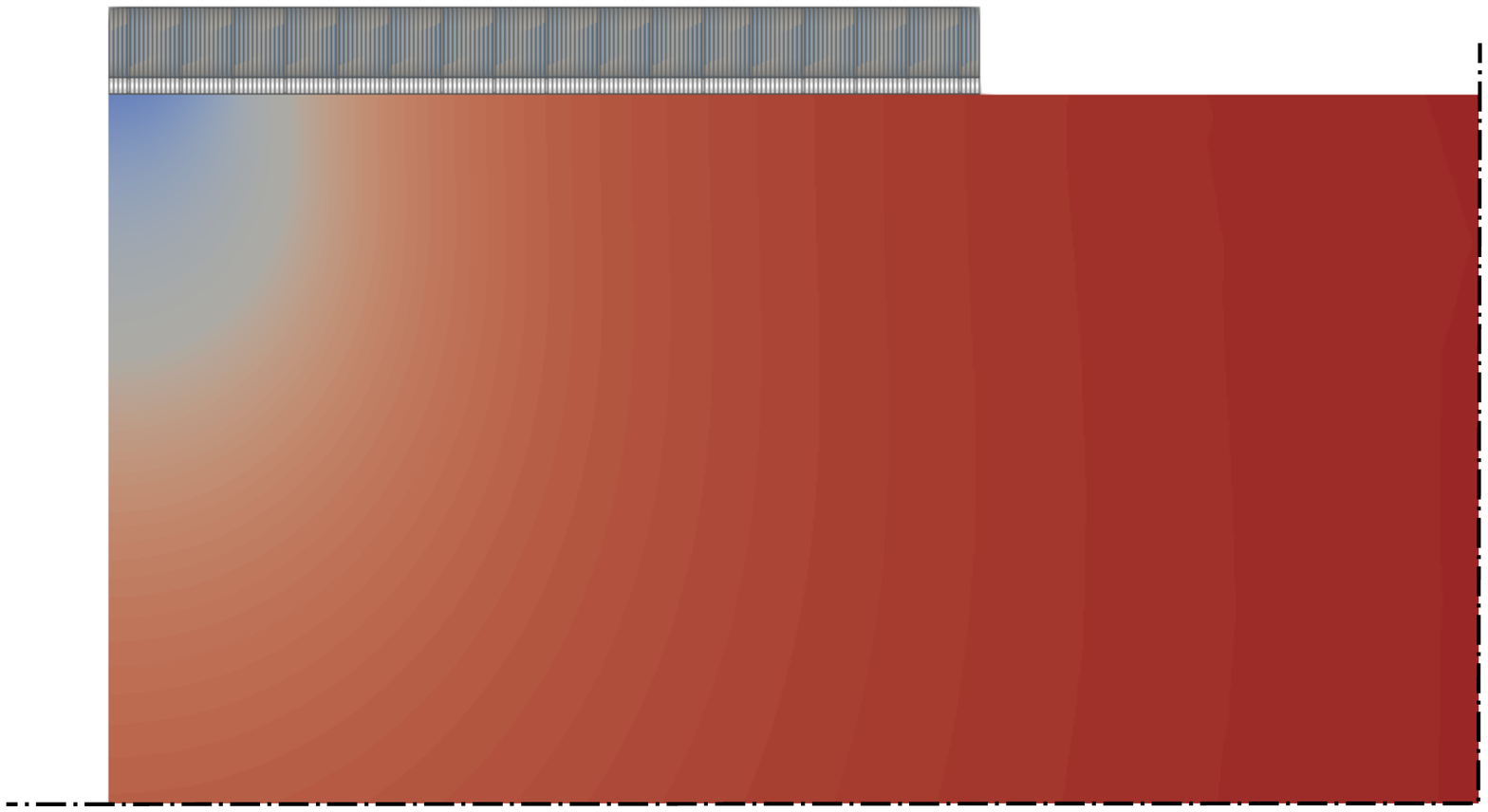}}
    \subfloat[$\bar{u}=5.4\times10^{-3}\millm$]{
    \includegraphics[width=0.4\linewidth]{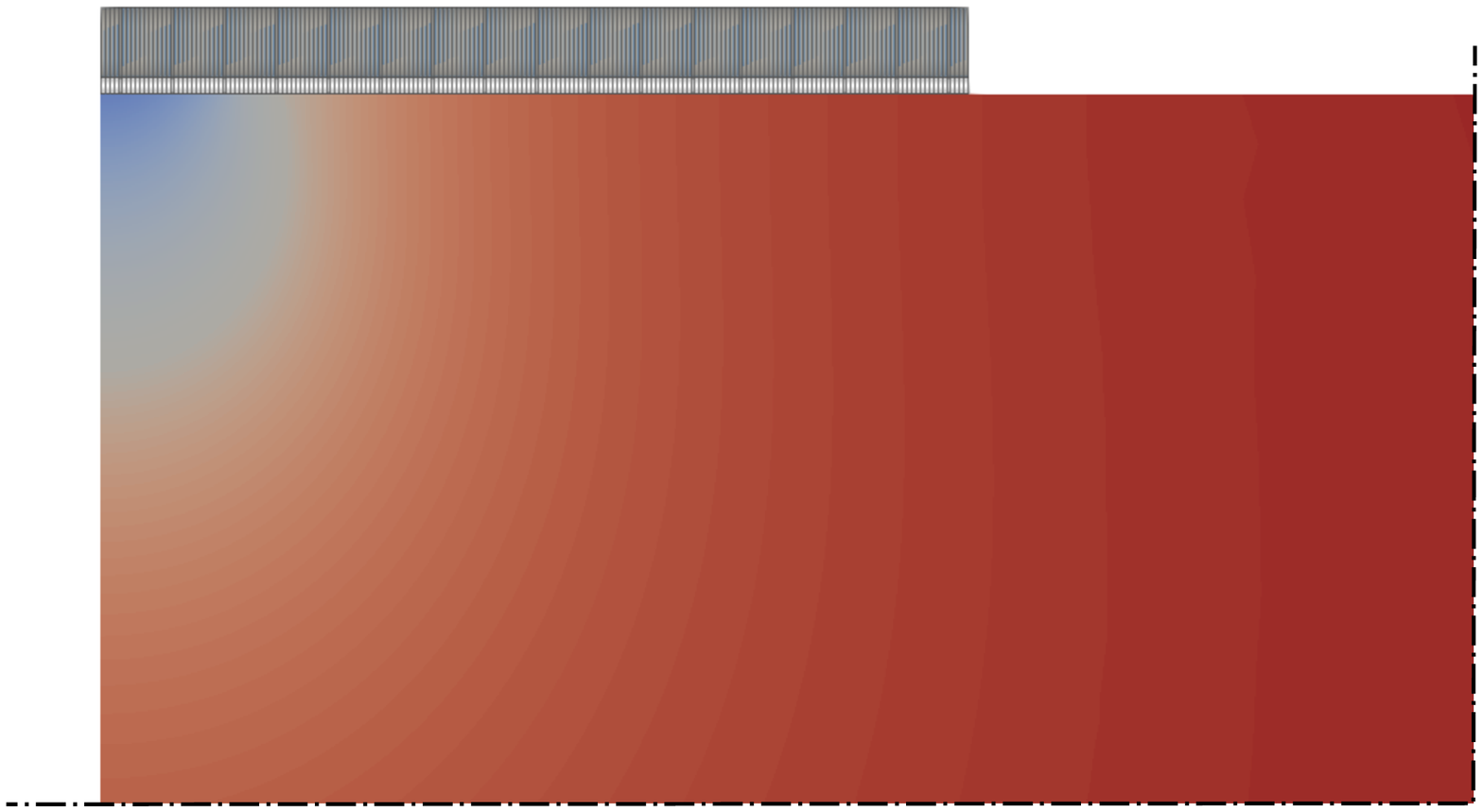}}\\
    \subfloat[$\bar{u}=6.5\times10^{-3}\millm$]{
    \includegraphics[width=0.4\linewidth]{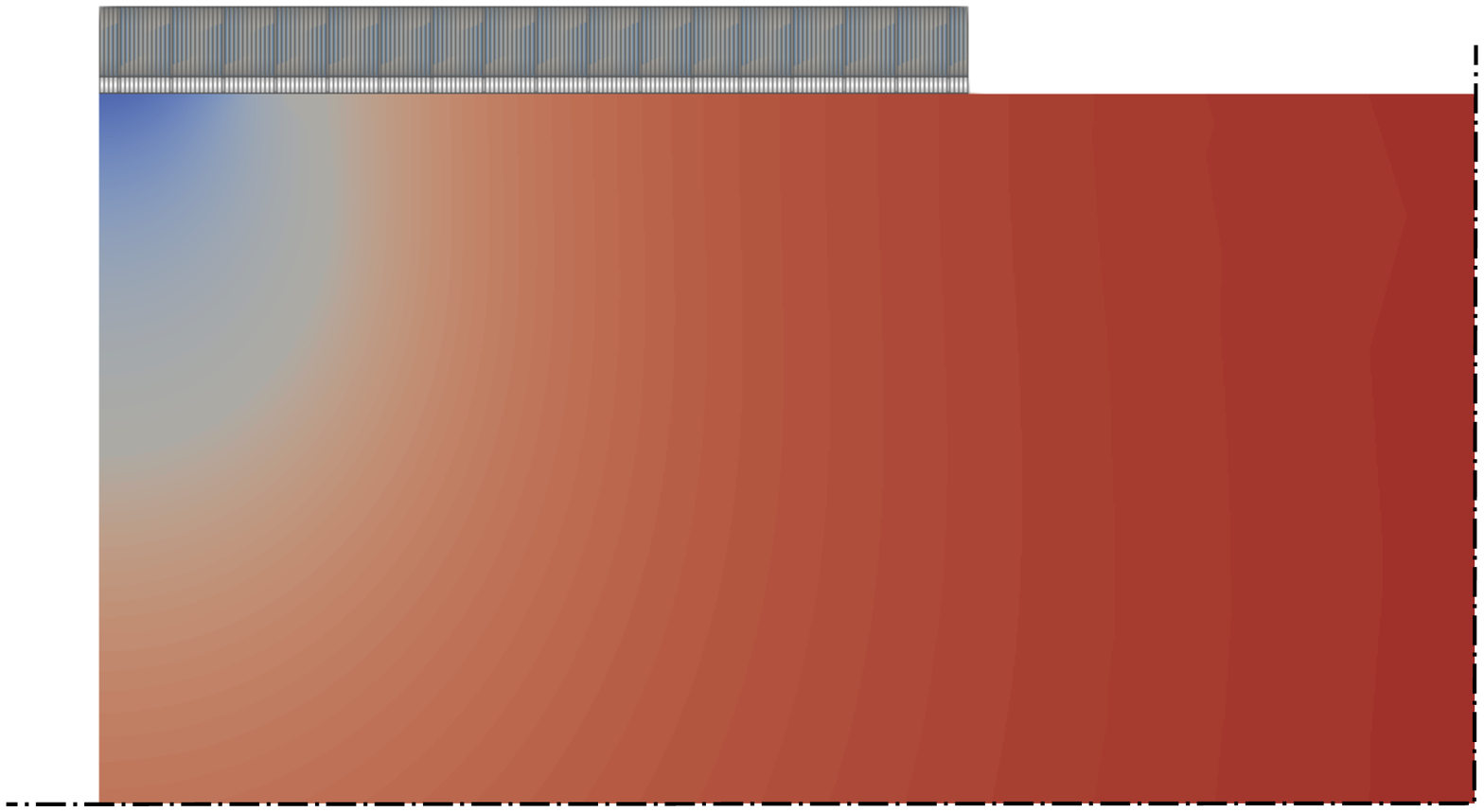}}
    \subfloat[$\bar{u}=8.4\times10^{-3}\millm$]{
    \includegraphics[width=0.4\linewidth]{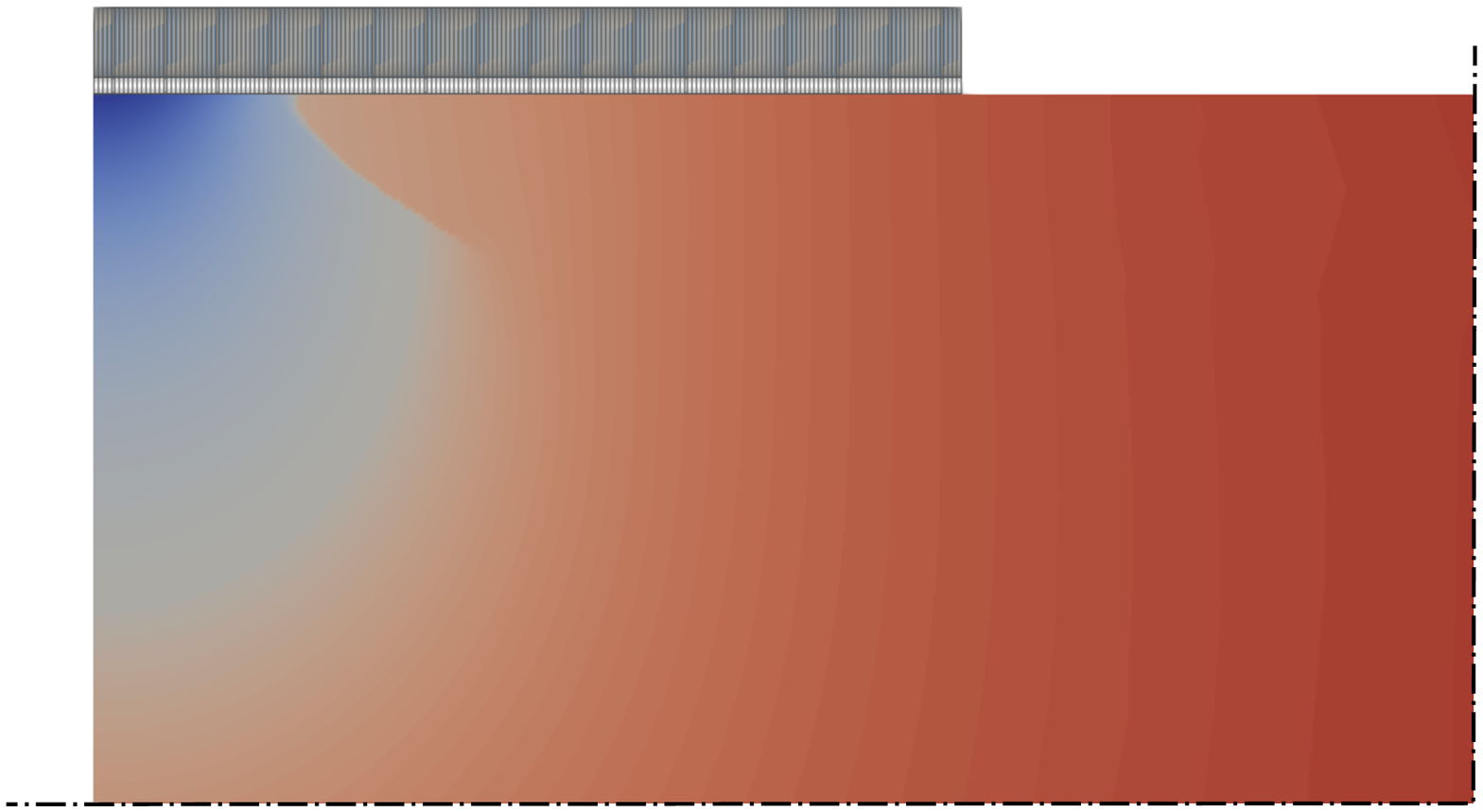}}\\
    \includegraphics[width=0.25\linewidth]{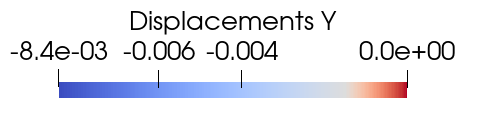}\\
    \caption{Snapshots of the contour plots of the vertical displacement component for the indentation test with a sphere of radius $R_s=1 \millm$. In (d), the displacement discontinuity due to fracture is evident from the jump from blue to red colors.}
    \label{fig:displ}
\end{figure}

The vertical displacement field plotted at different pseudo-time steps is shown in \fig\ref{fig:displ}, with the expected Hertzian distribution correctly reproduced (see also \citet{Bonari2020} for more details on the analysis of the stress field for this benchmark solution).

\begin{figure}[h!]
    \centering
    \includegraphics[width=0.7\linewidth]{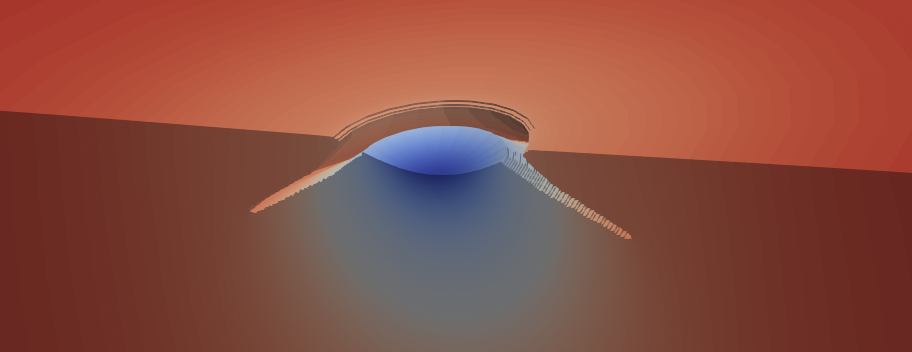}
    \caption{\textcolor{black}{3D post-process view of the 2D axisymmetric crack pattern where fractured finite elements with $\phi \cong 1$ have been removed, for ease of visualization, at a far field vertical displacement of $\bar{u}=0.01\millm$.}}
    \label{fig:crack_defo}
\end{figure}

\begin{figure}[h!]
    \centering
    \includegraphics[width=0.9\linewidth]    {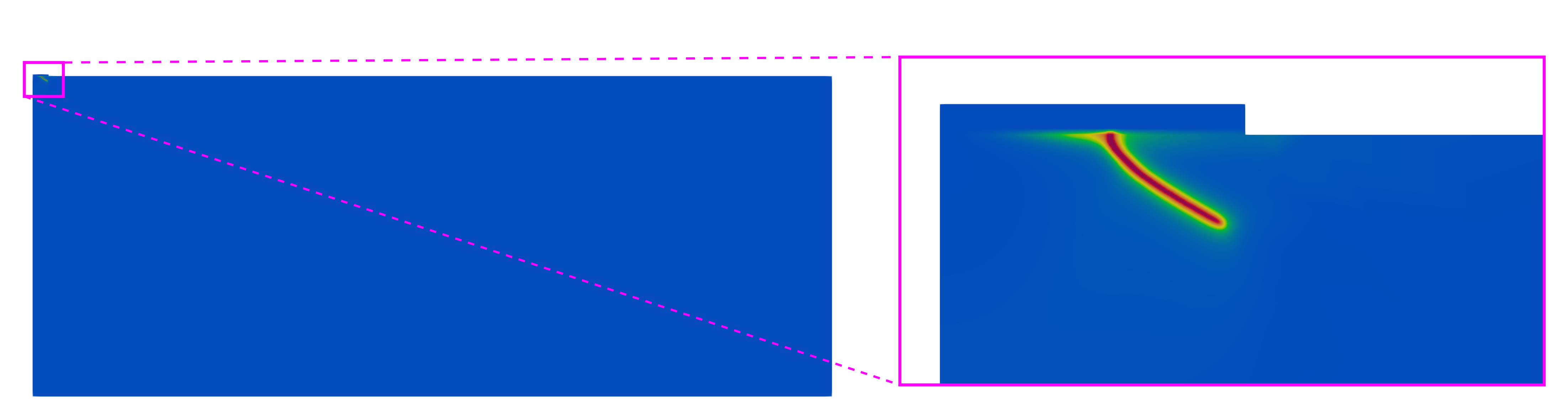}\\
    \includegraphics[width=0.2\linewidth]{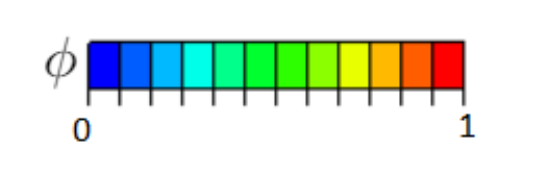}\\
    \caption{\textcolor{black}{Crack pattern due to the indentation with a sphere having $R=1\millm$ shown in the entire domain and in a magnified view.}}
    \label{fig:crack_entire_model}
\end{figure}

The crack develops with its typical conical shape, as shown in \fig\ref{fig:crack_defo}. \textcolor{black}{The contour plots presented in this section show only a portion of the entire domain to evaluate the crack pattern better. In order to clarify this aspect, the proportion between the dimension of the entire domain and the detail of the crack pattern is shown in \fig\ref{fig:crack_entire_model}.}

The fracture evolution at different pseudo-time steps is shown in \fig\ref{fig:crack_evol} for different values of the imposed far-field displacement $\bar{u}$. 

\FloatBarrier

\begin{figure}[h!]
    \centering
    \subfloat[]{
    \includegraphics[width=0.7\linewidth]{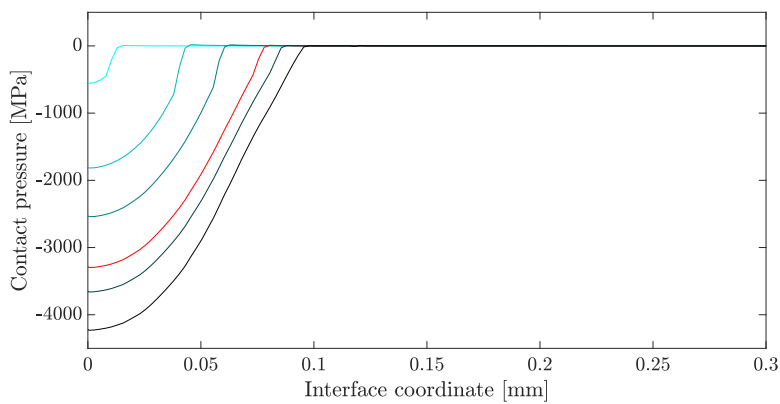}} \\
    \subfloat[]{
    \includegraphics[width=0.7\linewidth]{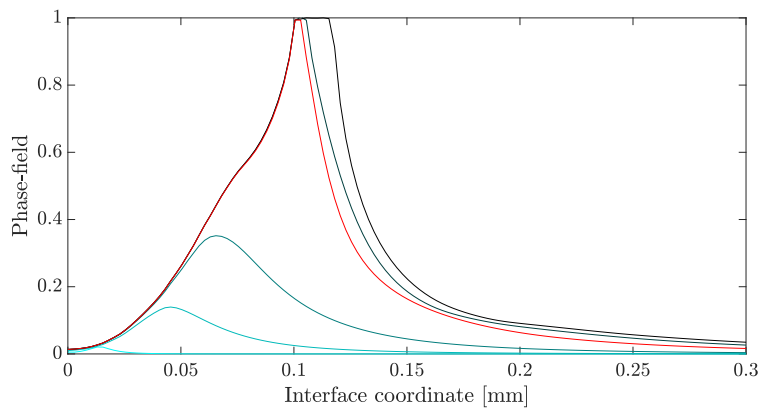}}
    \caption{Contact pressure in (a) and phase-field variable in (b) along the interface at different pseudo-time steps for the indentation test with $R_s=1\millm$. The red curves correspond to the variables at the point of crack nucleation.}
    \label{fig:cont_press}
\end{figure}

A quantitative analysis of the results shows that the interface contact pressure increases according to the Hertzian pressure distribution as shown in \fig\ref{fig:cont_press}(a), where the red curve corresponds to the crack onset. For the same pseudo-time steps, the phase-field variable evolution along the interface is shown in \fig\ref{fig:cont_press}(b). The contact pressure and the phase-field plots are used to evaluate, respectively, the contact radius at crack nucleation, $a_c$, and the ring crack radius, $r_0$. The contact radius at crack nucleation is defined by the coordinate where the contact pressure becomes vanishing after having been negative valued, for the red curve in \fig\ref{fig:cont_press}(a) that corresponds to the first point of the interface where the phase-field variable reaches unity. The coordinate of that point, which can be assessed from \fig\ref{fig:cont_press}(b), gives the ring crack radius, $r_0$. \textcolor{black}{Hence, the propagation of the crack outside the contact area can be clearly noted in \fig\ref{fig:cont_press} by comparing the two red curves which gives $r_0/a_c=1.28$.}

The effect of the sphere radius on the critical load causing crack nucleation has been investigated by considering five cases with a radius $R_s$ varying from $1$ to $7.5\millm$. The introduction of the interface finite elements allows changing the spherical indenter radius without changing the model geometry since the modified spherical geometry is embedded into the finite elements and analytically defined as a function of its radius.

\begin{figure}[h!]
    \centering
    \includegraphics[width=0.75\linewidth]{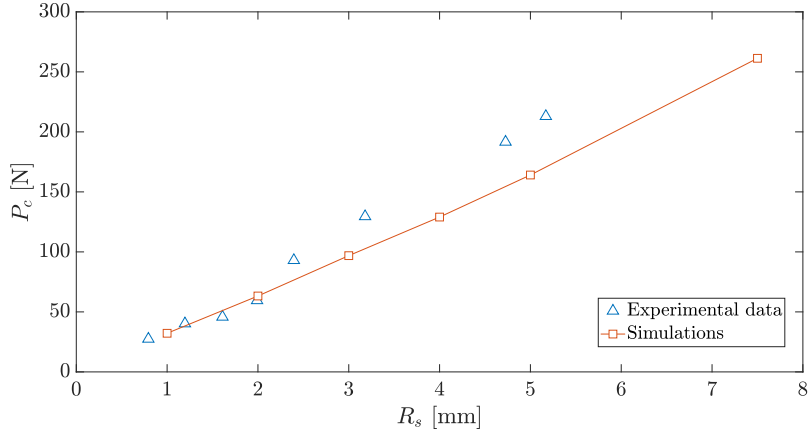}
    \caption{Critical load variation with respect to the sphere indenter radius and comparison with experimental data from \citet{Conrad1979} for the case of as-received (smooth) specimens.}
    \label{fig:crit_load}
\end{figure}

The results of the simulations are collected in \fig\ref{fig:crit_load} where current simulations are compared with the values obtained in the experimental campaign in \citep{Conrad1979} for smooth spheres. This graph shows a very satisfactory agreement between the experimental and numerical data.

\begin{figure}[h!]
    \centering
    \subfloat[\label{fig:sim_ring_ratio}  
    \textcolor{black}{Simulations results compared with the experimental data in \citep{Jyh-Woei1993} on as-received glass substrates.}]{
    \includegraphics[width=0.4\linewidth]{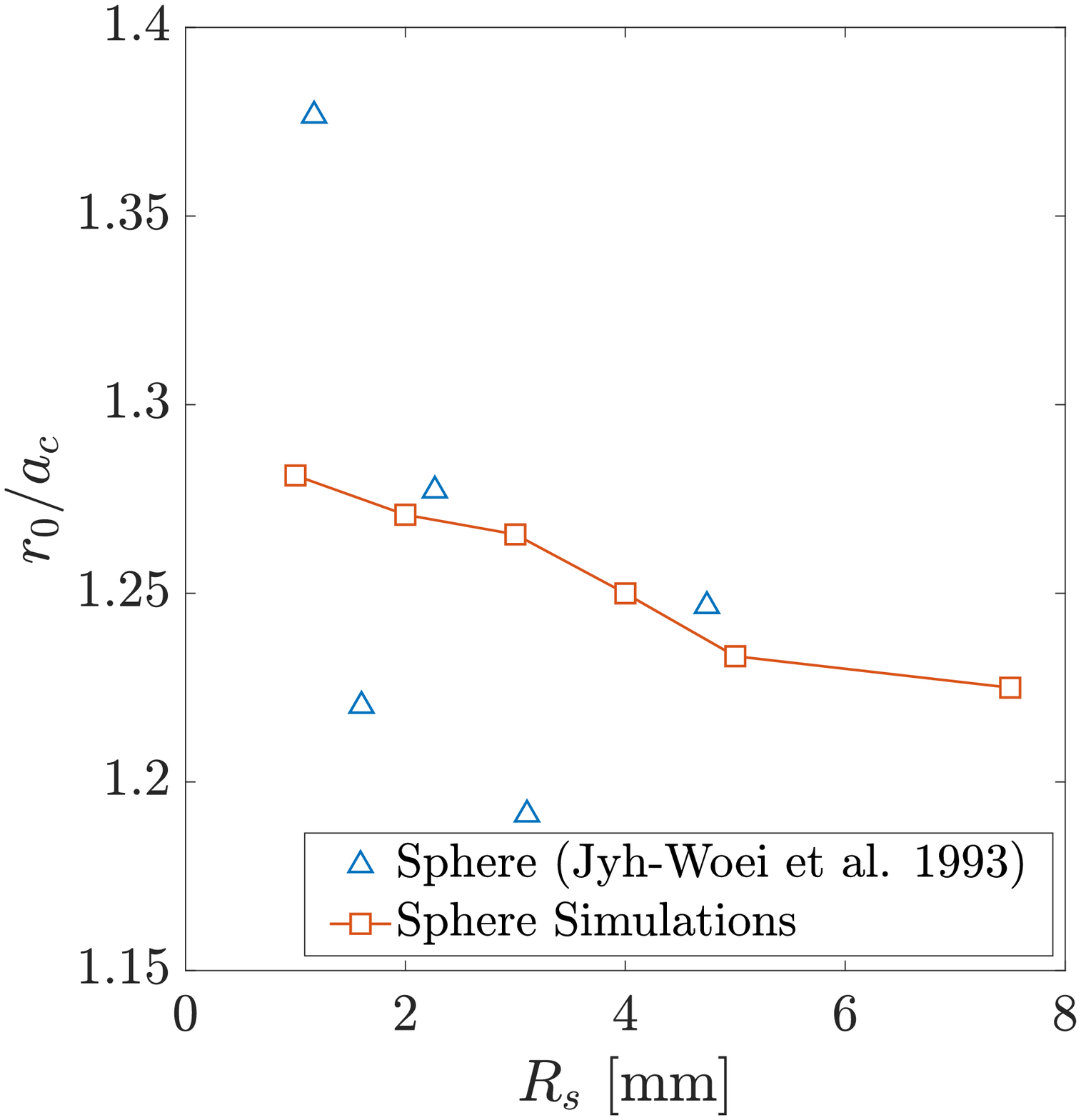}}
    \qquad
    \subfloat[\label{fig:moug_ring_ratio} 
    \textcolor{black}{Indentation of flat-ended cylinders and spheres on abraded glass substrates with 1000 grit silicon carbide paper in \citep{Mouginot1985}.}]{
    \includegraphics[width=0.4\linewidth]{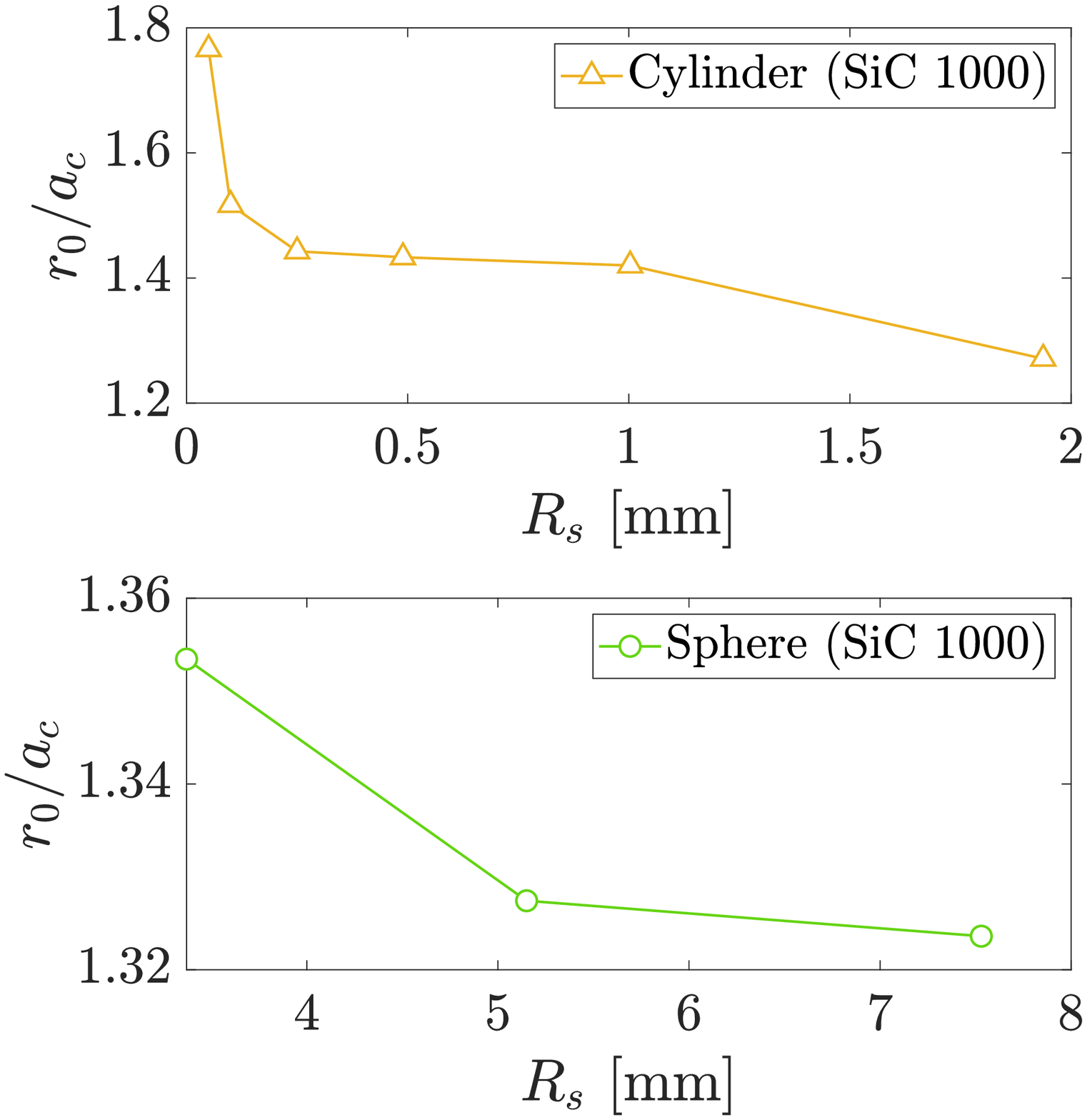}}
    \caption{\textcolor{black}{Ratio between the crack radius and the critical contact radius \textit{vs} the indenter radius. The experimental data in (b) concern indentation tests on glass specimens with different mechanical and geometrical properties.}}
    \label{fig:ratio_ring_contact_rad}
\end{figure}

The ring crack radius $r_0$ is always greater than the critical contact radius $a_c$, consistently with results in \citep{Mouginot1985,Conrad1979,Jyh-Woei1993}, see the plots in \fig\ref{fig:ratio_ring_contact_rad} where the ratio $r_0/a_c$ is related to the indenter radius, $R_s$. 

\textcolor{black}{Since no experimental data are available in \citep{Conrad1979}, the simulations results have been compared in \fig\ref{fig:sim_ring_ratio} against the indentation tests presented in \citep{Jyh-Woei1993} where the authors used the same type of glass and test geometry for the specimens. In this case, the scatter in the experimental data does not allow to confirm the obtained numerical trend. However, although a one-to-one comparison is not possible, the predicted trend is consistent with that found in the experiments in \citep{Mouginot1985} for flat-ended cylindrical punch (\fig\ref{fig:moug_ring_ratio}, upper panel) and for spherical indenters on glass specimens abraded with 1000 grit silicon carbide paper (\fig\ref{fig:moug_ring_ratio}, lower panel).}

\textcolor{black}{Although those experimental trends have been obtained with glass specimens of different sizes ($50 \millm \times 50 \millm \times 25.4\millm$) and material properties (borosilicate glass with $E= 80\gigap$, $\nu=0.22$ ) than those used in the simulations in \fig\ref{fig:sim_ring_ratio}, the trends are fully consistent and confirm that the ratio $r_0/a_c$ decreases by increasing the indenter radius.}

\begin{figure}[h!]
    \centering
    \subfloat[\label{fig:conrad_ratio_ring_inverse}
    \textcolor{black}{Results of the numerical simulations compared with the experimental data in \cite{Conrad1979}.}]{
    \includegraphics[width=0.4\linewidth]{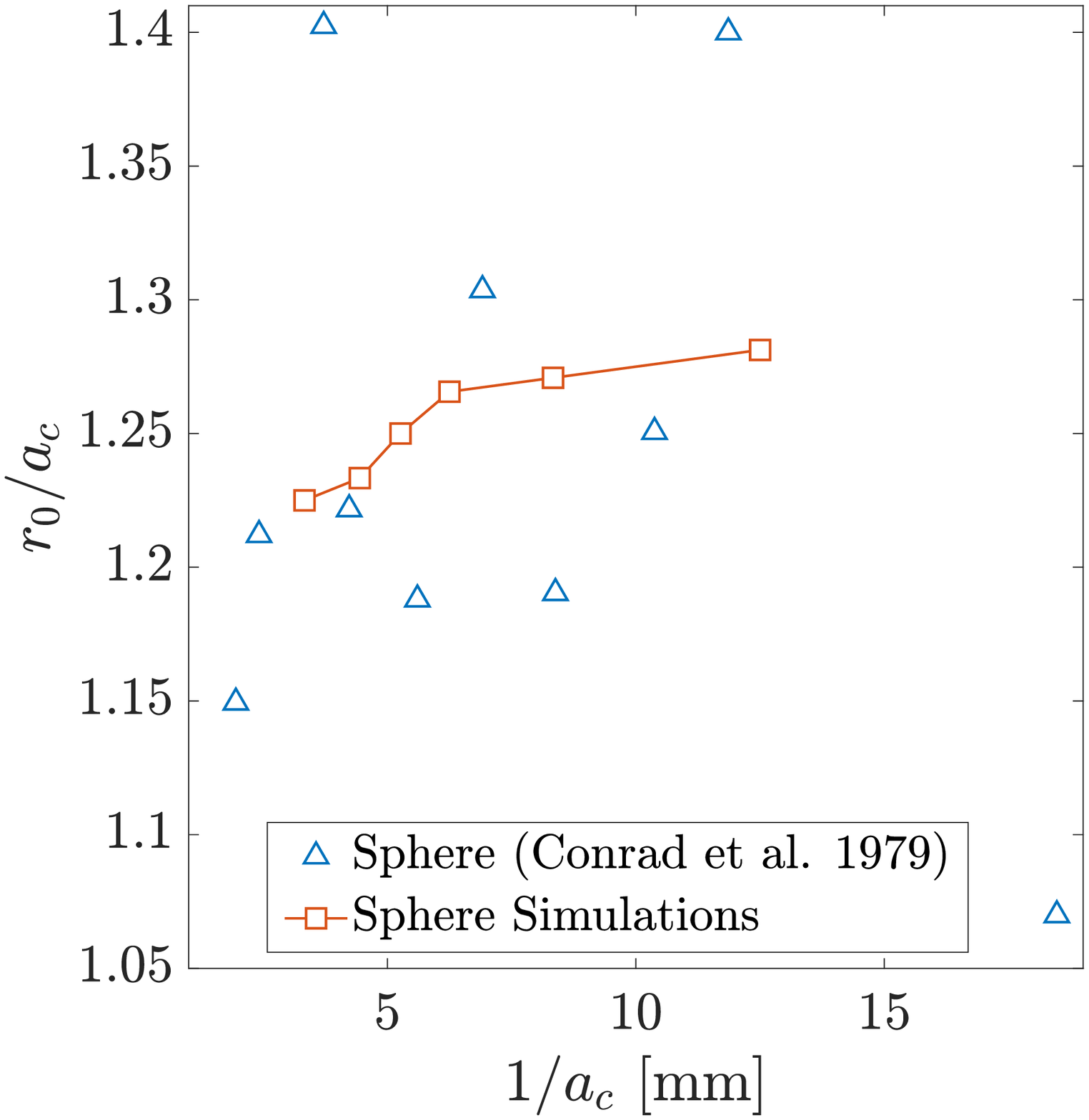}}
    \qquad
    \subfloat[\label{fig:moug_ratio_ring_inverse}
    \textcolor{black}{Indentation of flat-ended cylinders and spheres on glass abraded with 1000 grit SiC paper or $7\mum$ diamond paper in \cite{Mouginot1985}}]{
    \includegraphics[width=0.4\linewidth]{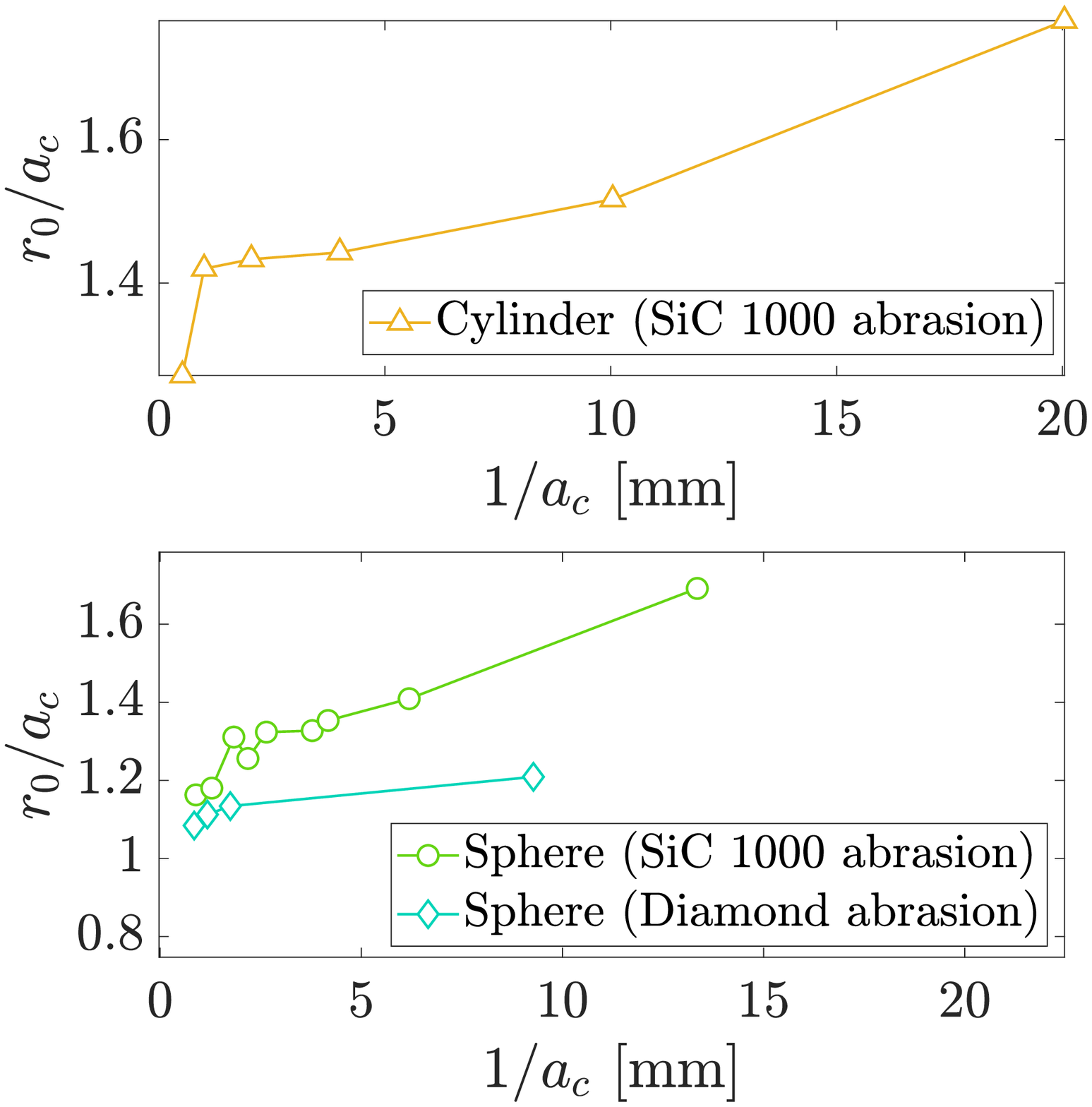}}
    \caption{\textcolor{black}{Ratio between the crack radius and the critical contact radius \textit{vs} the inverse of the contact radius.}}
    \label{fig:ratio_ring_contact_rad_inverse}
\end{figure}

\textcolor{black}{The variation of the ratio $r_0/a_c$  has been related to the inverse of the contact radius $1/a_c$ in \fig\ref{fig:conrad_ratio_ring_inverse} and compared with the experimental data in \citep{Conrad1979} which show again a very high scatter. For this reason, the trend resulting from the numerical model has been compared also with that resulting from the experimental tests in \citep{Mouginot1985} for different types of indenter, see \fig\ref{fig:moug_ratio_ring_inverse}. It has to be noted that in the case of a flat-end cylindrical punch, the radius of the indenter $R_s$ always coincides with the contact radius, which does not happen in the case of non-conformal contacts, as explained in the introduction. }

\textcolor{black}{The data on the spherical indenters in \fig\ref{fig:moug_ring_ratio}, as well as the investigations in \citep{Conrad1979, Jyh-Woei1993}, show that the surface treatment (abrasion with 1000 grit SiC paper, or with $7 \mum$ diamond paste) influences the indentation tests and increases the $r_0/a_c$ ratio. This latter aspect will be treated in the following paragraphs dealing with the case of rough indenters.}

\FloatBarrier

To further exploit the advantage of the present approach, which enables to embed any indenter profile along a nominally flat interface and efficiently solve the contact problem, we now consider the effect of surface roughness on cone indentation fracture, using the same axial symmetric finite element model for the smooth sphere, but modifying the embedded indenter profile. 

\textcolor{black}{Generally, a real rough surface does not present axial symmetry; however, a 3D indentation model requires a high computational effort compared to the axisymmetric configuration.}
\textcolor{black}{For example, the 2D axisymmetric model of the smooth spherical indentation test presented at the beginning of this section requires a computational time of around $30$ min to solve 100 pseudo-time steps up to an imposed displacement of $0.01\millm$. In the case of a 3D simulation of the same problem, because of the drastic increase of the degrees of freedom, parallel computing facilities are required: with 20 cores, the CPU time for a 3D FE simulation becomes comparable with the CPU time for a single CPU calculation of the 2D axisymmetric problem. Moreover, for the 3D simulations of a rough sphere, the CPU time increases up to 5 h, even with parallel computing, because of the more complex damage pattern, as shown later. For this reason, the 2D axisymmetric configuration has been chosen for the simulations in this work. To the best of the authors' knowledge, even the 2D axisymmetric solutions represent the first attempt to investigate the effect of roughness for this kind of complex nonlinear coupled problem involving contact and fracture.} 

\begin{figure}[h!]
    \centering
    \subfloat[Profile obtained with RMD, with $R_z=1.3\,\si{\mu m}$ and sampling $2.5 \mum$.]{
    \includegraphics[width=0.6\linewidth]{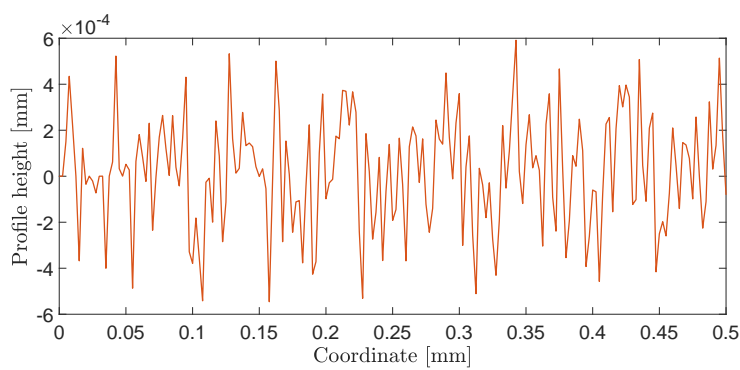}} \\
    \subfloat[\textcolor{black}{Final profile of the indenter embedded into the interface.}]{
    \includegraphics[width=0.65\linewidth]{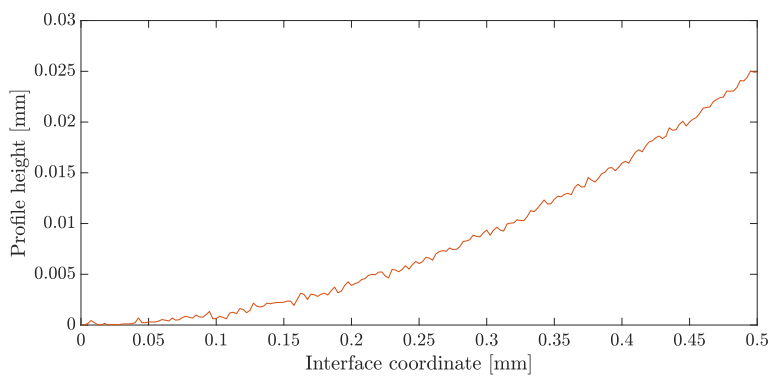}}
    \caption{Profile of the rough spherical indenter (b) embedded into the MPJR interface finite elements, obtained by superimposing the rough profile (a) to a sphere of radius $R_s=5\millm$.}
    \label{fig:rough_prof}
\end{figure}

For this purpose, one rough surface has been generated using the Random Midpoint Displacement (RMD) algorithm already widely exploited in the contact mechanics literature to simulate realistic surface roughness \citep{Paggi2010}, with a spatial resolution of $2.5 \mum$, and fractal dimension $D=2.1$. From the numerically generated rough surface, one profile has been extracted and superimposed to the spherical shape of the smooth indenter (see the profile in \fig\ref{fig:rough_prof}(a) and the resulting shape of the indenter in \fig\ref{fig:rough_prof}(b). 

\textcolor{black}{In the experimental data in \citep{Mouginot1985, Jyh-Woei1993, Conrad1979}, the authors did not directly measure the specimens' roughness due to the abrasion process; however, they reported the presence of defects on the surface of a few microns after the treatment with grit papers or diamond paste. The rough profile chosen for the simulation has the statistical parameter $R_z$ equal to $1.3\mum$, which measures the average peak-to-valley distance of the profile.}

\begin{figure}[h]
    \centering
    \subfloat[$\bar{u}=9.2\times10^{-3}\millm$]{
    \includegraphics[width=0.48\linewidth]{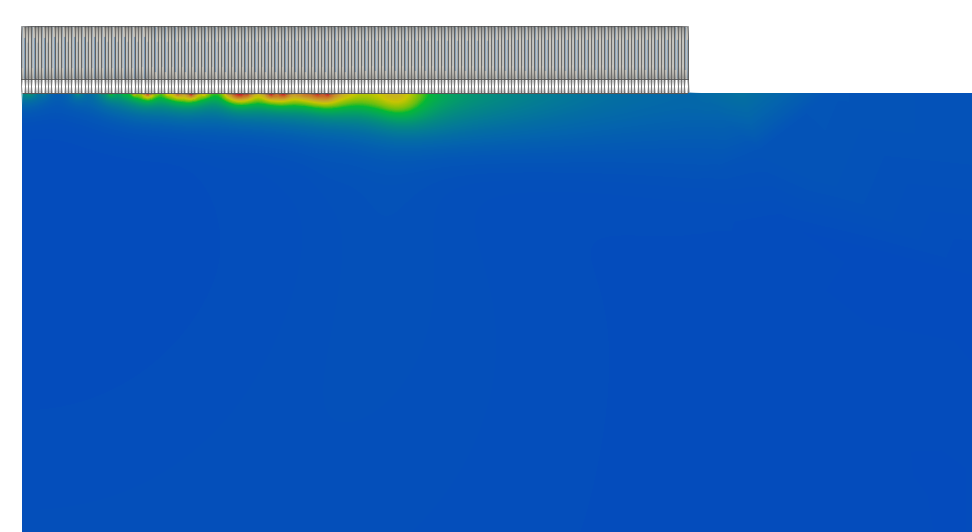}}
    \subfloat[$\bar{u}=9.3\times10^{-3}\millm$]{\includegraphics[width=0.48\linewidth]{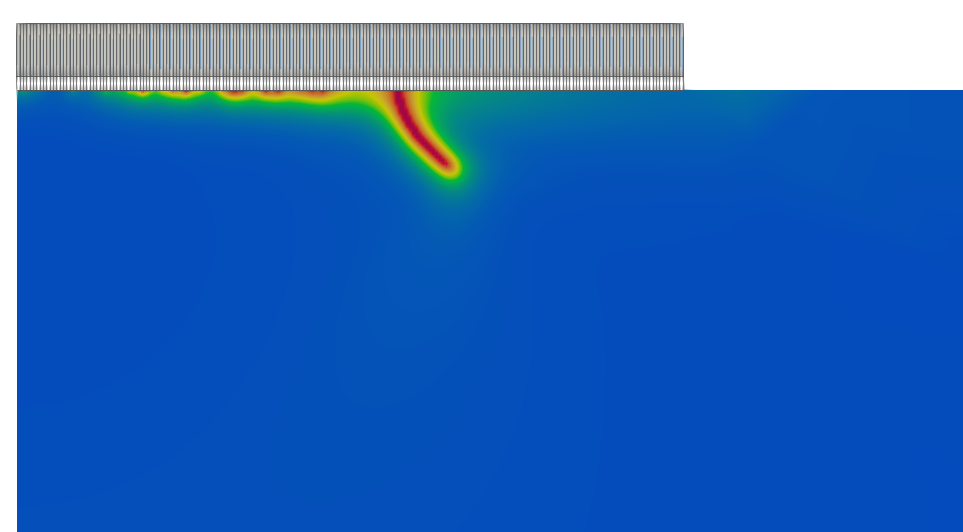}}\\
    \subfloat[$\bar{u}=9.6\times10^{-3}\millm$]{\includegraphics[width=0.48\linewidth]{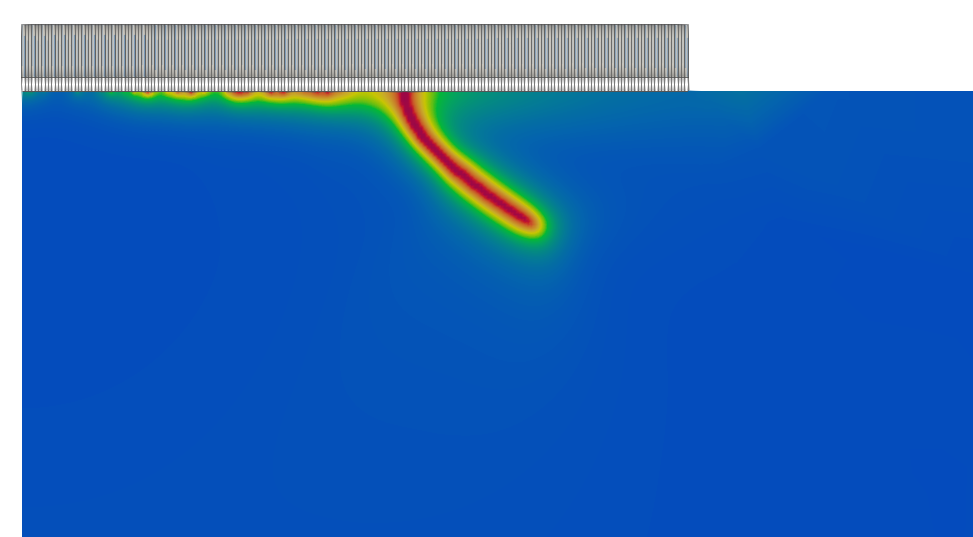}}
    \subfloat[$\bar{u}=0.01\millm$]{\includegraphics[width=0.48\linewidth]{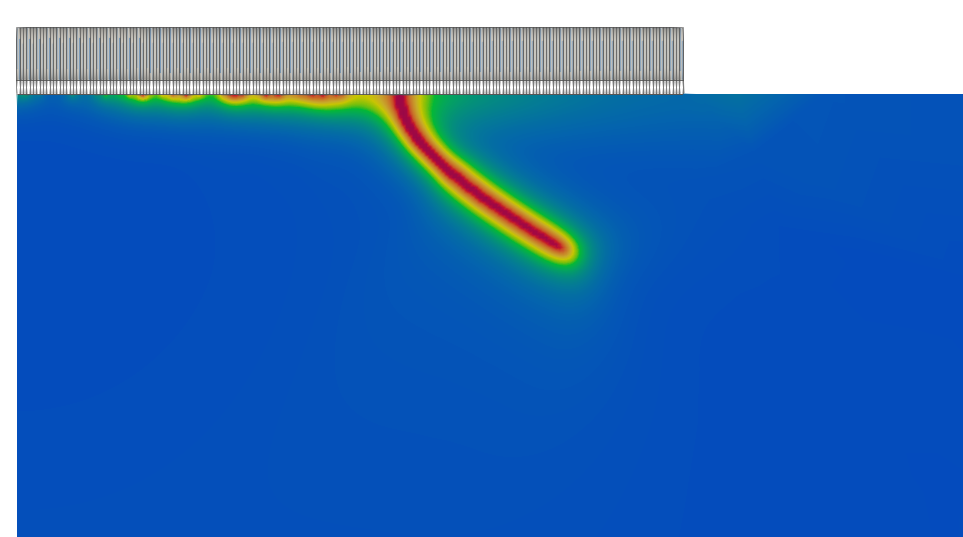}}    \\
    \includegraphics[width=0.2\linewidth]{legend_horizontal.pdf}
    \caption{Evolution of damage and the main crack for the indentation of glass with a rough spherical indenter with $R_z=1.3\,\si{\mu m}$.}
    \label{fig:crack_evol_rough}
\end{figure}

The fracture evolution for the case with rough indenter radius $R_s=5\millm$ is detailed in \fig\ref{fig:crack_evol_rough}. The crack propagates with a conical shape, as in the case of a smooth spherical indenter. However, the presence of roughness causes local damage at different points on the glass surface before the main crack propagates. 

\begin{figure}[h]
    \centering
    \subfloat[]{
    \includegraphics[width=0.78\linewidth]{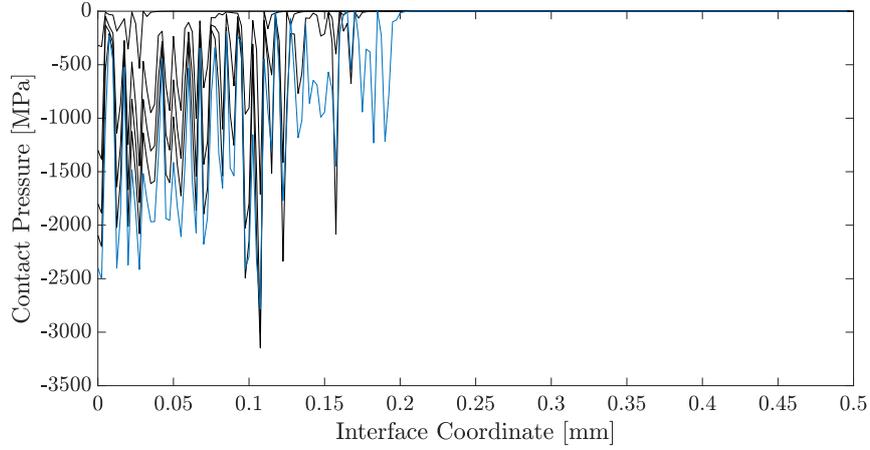}}\\
    \subfloat[]{
    \includegraphics[width=0.78\linewidth]{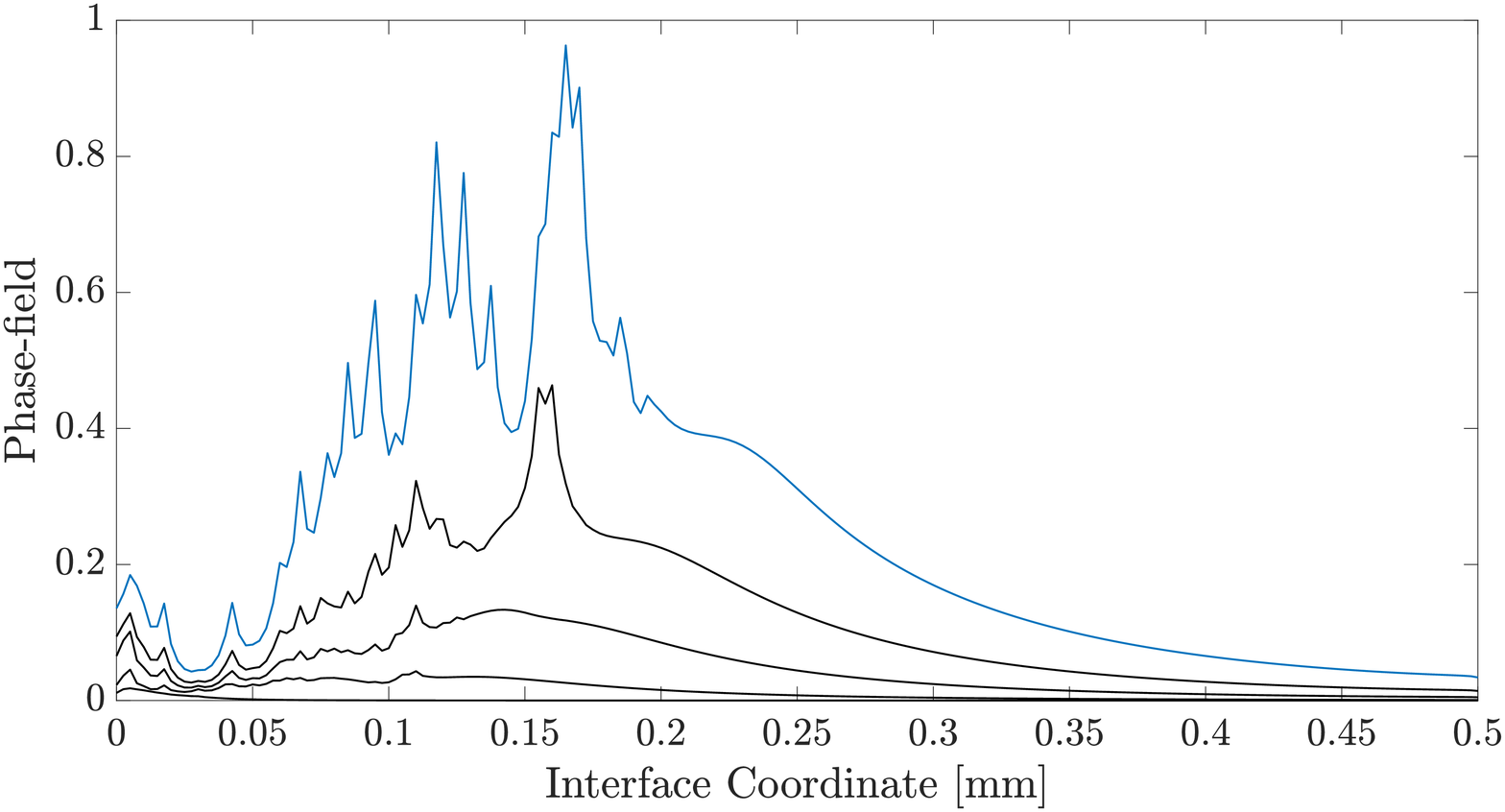}}\\
    \caption{\textcolor{black}{Rough spherical indenter with $R_z=1.3\,\si{\mu m}$: (a) contact pressure; (b) phase-field variable for different pseudo-time steps up to the occurrence of the first point along the interface with $\phi \approx 1$ (blue curve, corresponding to a displacement $\bar{u}=9.2 \times 10^{-3}\millm$).}}
    \label{fig:cp_pf_1}
\end{figure}

\begin{figure}[h]
    \centering    
    \subfloat[]{
    \includegraphics[width=0.8\linewidth]{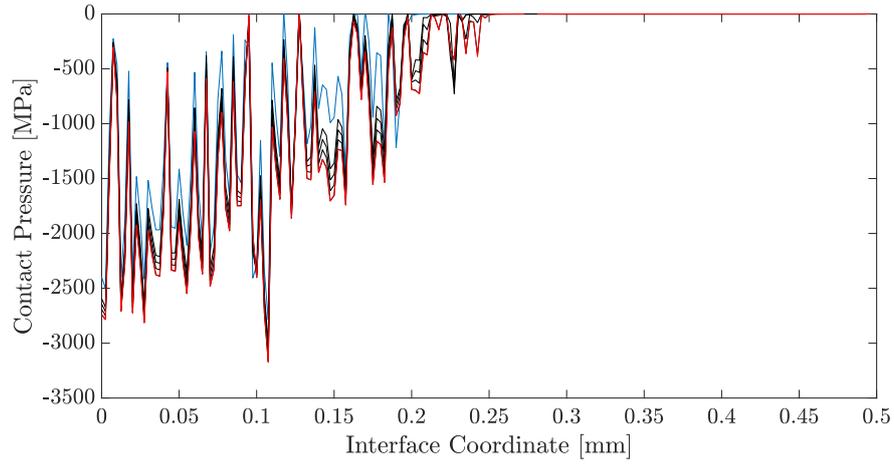}}\\
    \subfloat[]{
    \includegraphics[width=0.8\linewidth]{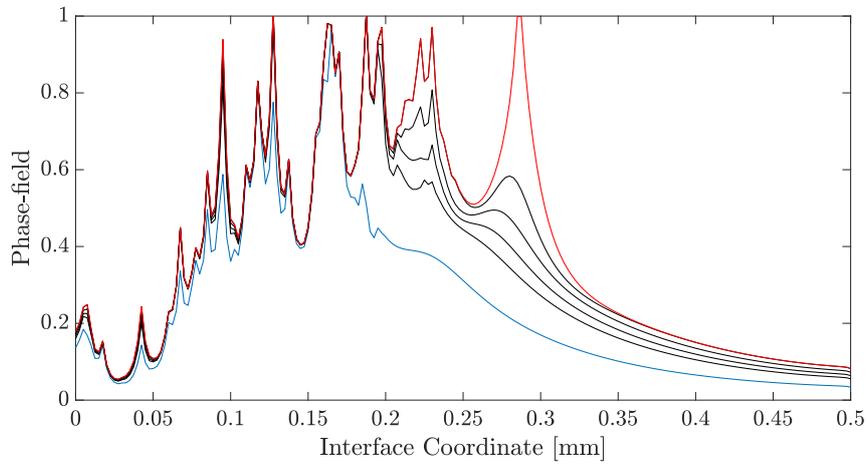}}
    \caption{\textcolor{black}{Rough spherical indenter with $R_z=1.3\,\si{\mu m}$: (a) contact pressure; (b) phase-field variable for different pseudo-time steps from the occurrence of the first point along the interface with $\phi \approx 1$ (blue curve), up to the situation corresponding to crack propagation for a higher displacement equal to $\bar{u}=9.3 \times 10^{-3}\millm$ (red curve).}}
    \label{fig:cp_pf_2}
\end{figure}

\begin{figure}[h]
    \centering
    \subfloat[]{
    \includegraphics[width=0.8\linewidth]{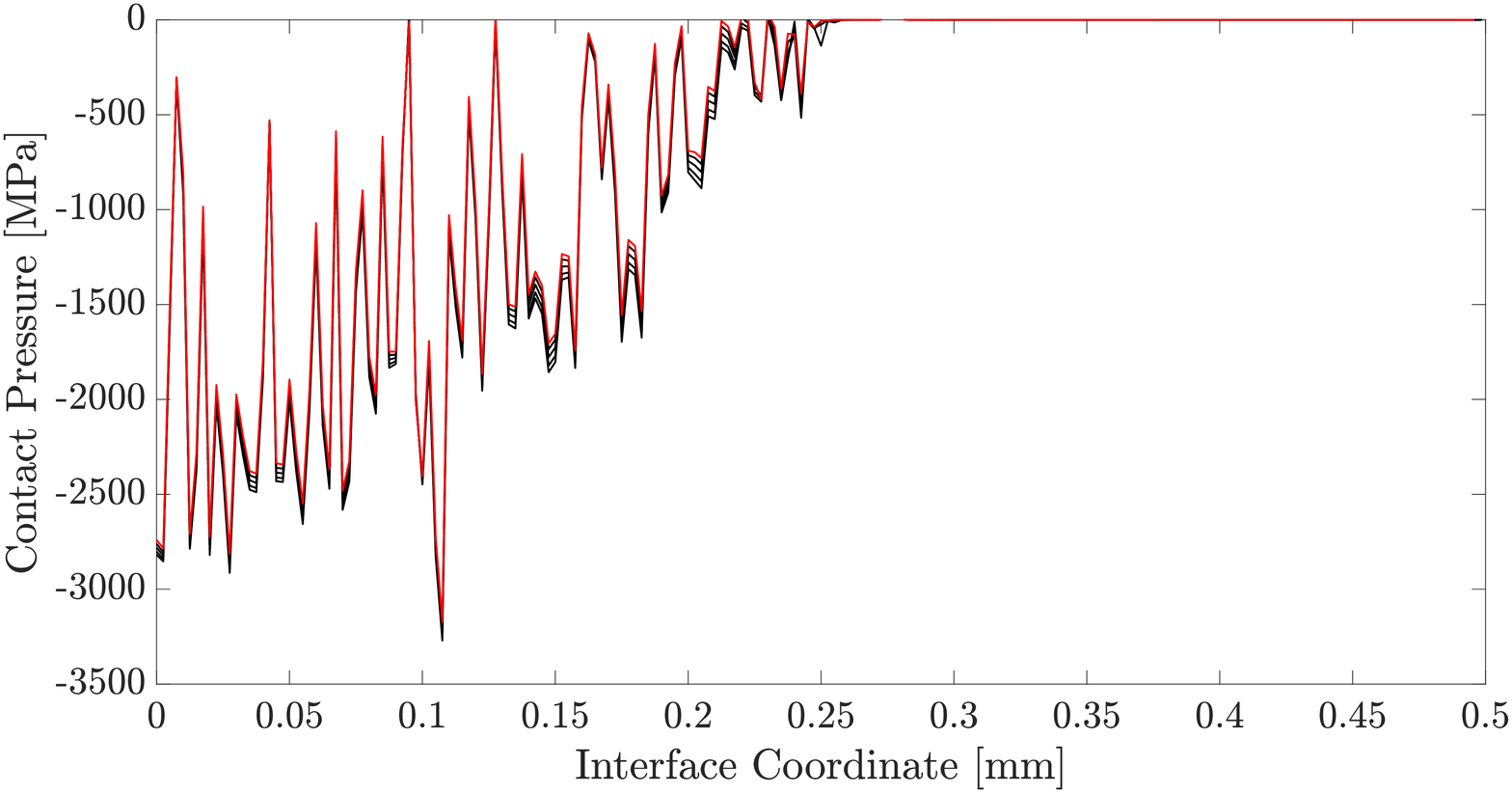}}\\
    \subfloat[]{
    \includegraphics[width=0.8\linewidth]{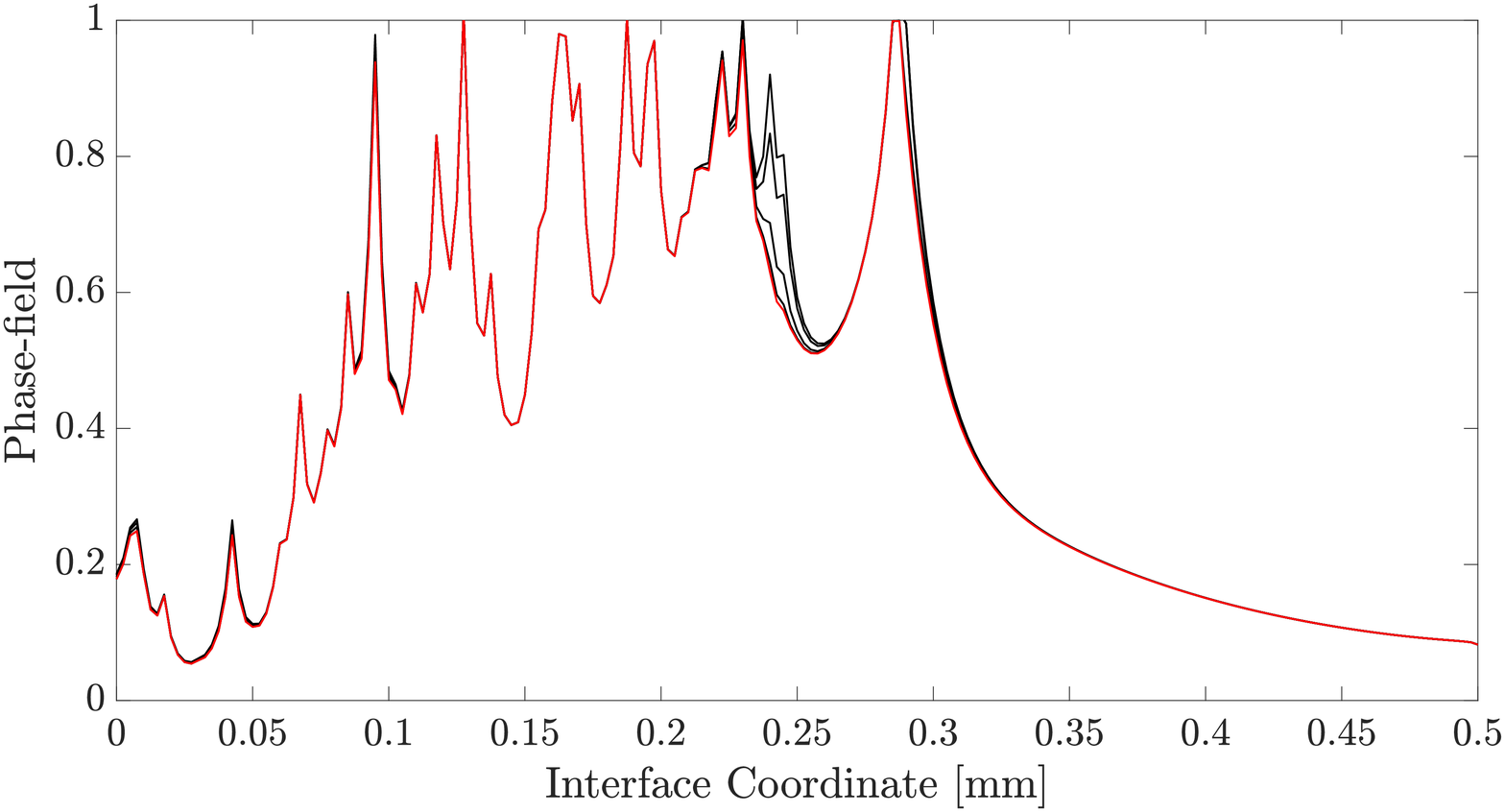}}
    \caption{\textcolor{black}{Rough spherical indenter with $R_z=1.3\,\si{\mu m}$: (a) contact pressure; (b) phase-field variable for different pseudo-time steps from the occurrence of crack propagation (red curve) onward, up to an imposed displacement equal to $\bar{u}=0.01\millm$.}}
    \label{fig:cp_pf_3}
\end{figure}

\textcolor{black}{The local damage can also be seen by looking at the phase-field variable plotted along the interface in \figs\ref{fig:cp_pf_1}, \ref{fig:cp_pf_2} and \ref{fig:cp_pf_3} together with the contact pressure at the interface.
In \fig\ref{fig:cp_pf_1}, the two variables have been plotted at different time steps until the phase-field variable is $\approx 1$ at the interface coordinate $x=0.175\millm$, see the blue curve corresponding to the imposed displacement ${\bar{u}=9.2\times10^{-3}\millm}$. 
The main crack does not propagate from that point but from the last peak of the phase-field red curve in \fig\ref{fig:cp_pf_2} at $x=0.285\millm$, representing the ring radius $r_0$. This point is again outside the contact area, as in the case of the smooth indenter, since the last point in contact has a radial coordinate $x=0.2625\millm$.
No further damage along the interface can be seen after the main crack propagates, as shown in \fig\ref{fig:cp_pf_3}.}

\begin{figure}[h]
    \centering
    \subfloat[Contact Pressure comparison.]{
    \includegraphics[width=0.45\linewidth] {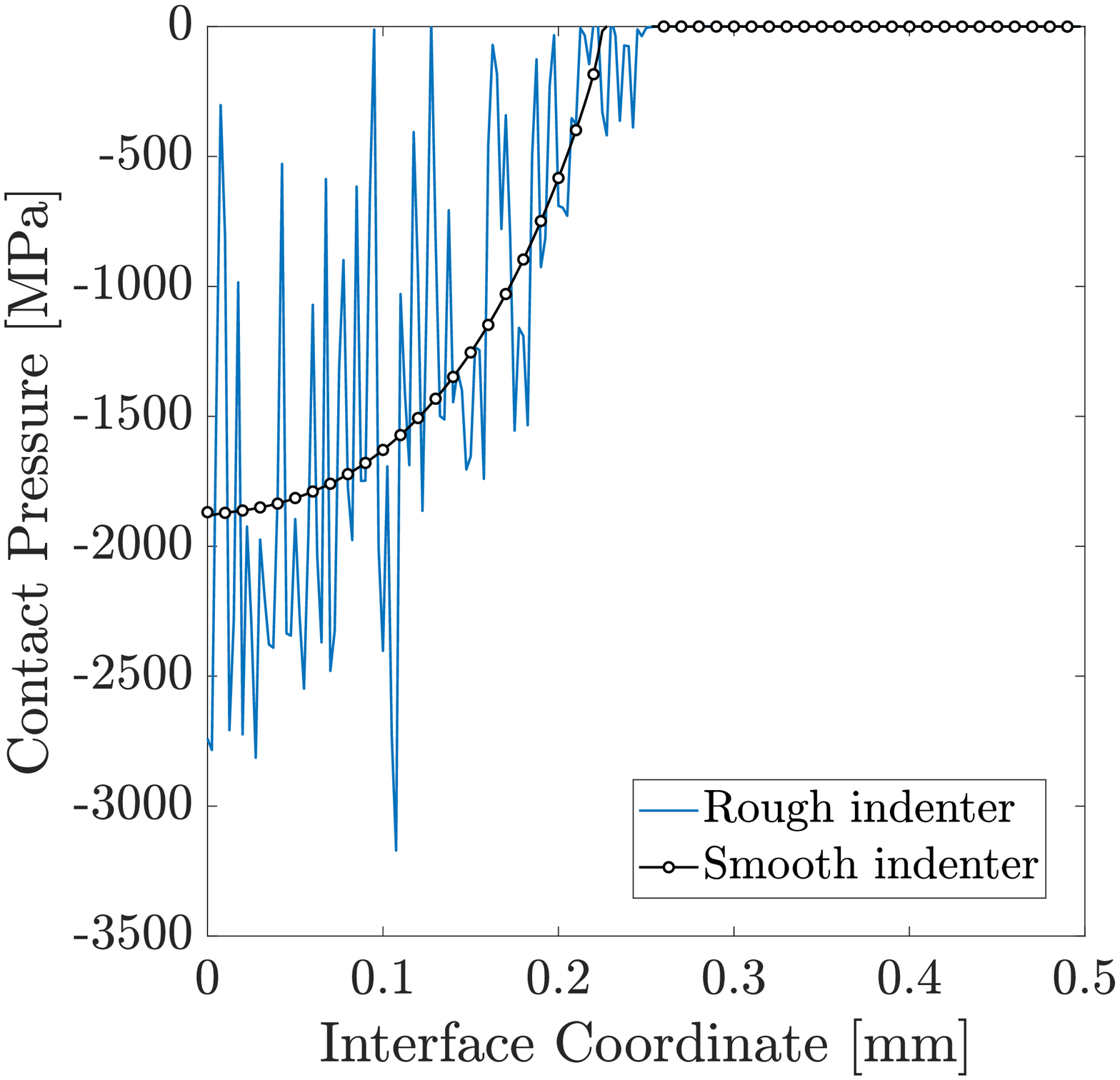}} \qquad
    \subfloat[Phase-field variable comparison.]{\includegraphics[width=0.45\linewidth]{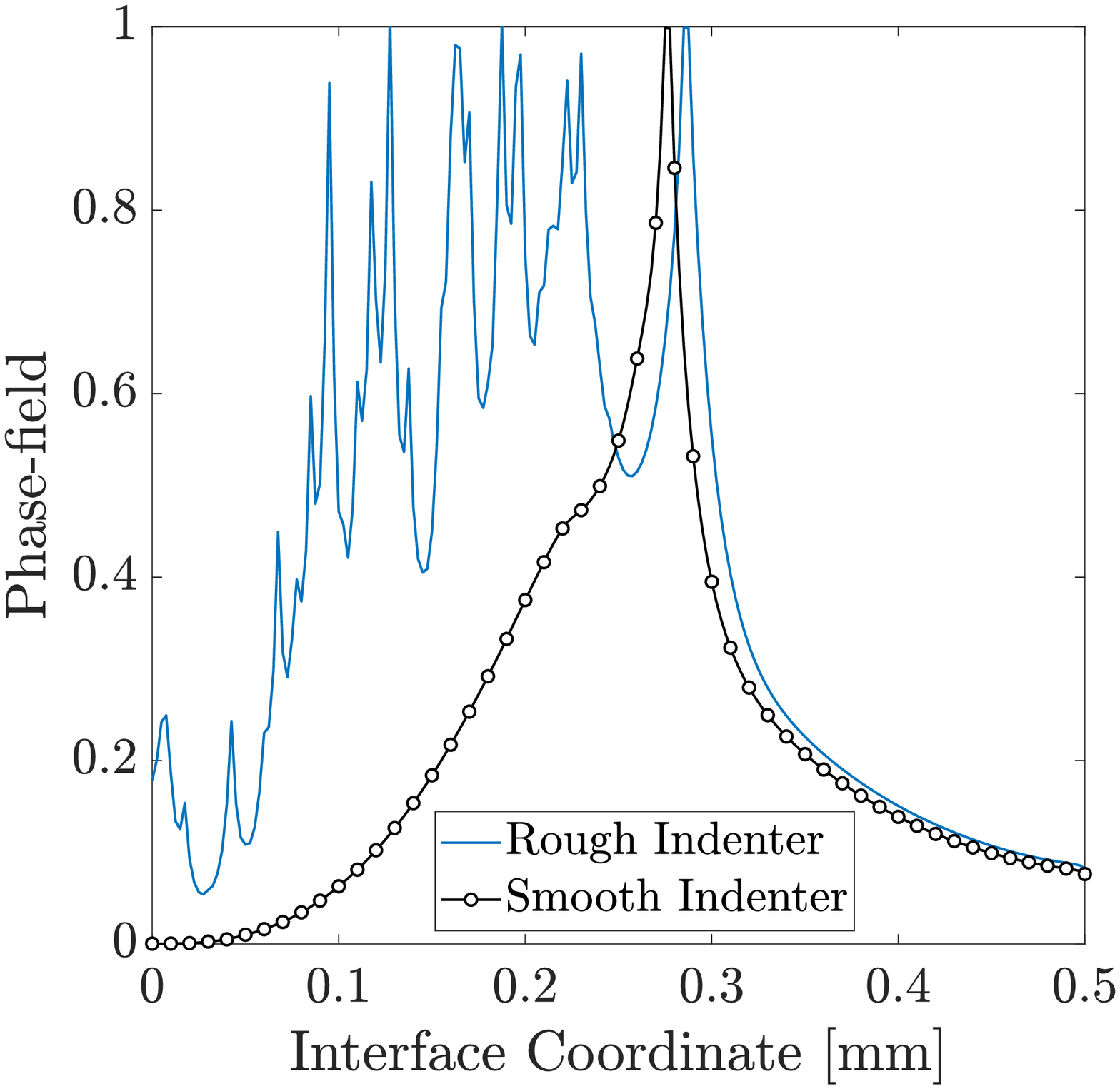}}
    \caption{\textcolor{black}{Comparison between smooth and rough spherical indenters $(R_s=5\millm)$ at the onset of crack nucleation, respectively at the imposed far-field displacements of $\bar{u}=0.92\millm$ and $\bar{u}=0.93\millm$ in the two cases.}}
    \label{fig:smooth_rough}
\end{figure}

\textcolor{black}{The case of the rough indenter has been compared with the smooth case, considering the same spherical radius $R_s=5\millm$, in terms of contact pressure and phase-field along the interface at the onset of main crack propagation, which happens at an imposed displacement equal to $\bar{u}=0.92\millm$ for the smooth sphere and $\bar{u}=0.93\millm$ for the rough indenter. The comparison shown in \fig\ref{fig:smooth_rough} highlights that roughness induces stress concentrations and damage at the points in contact, which do not occur for smooth profiles.}

\begin{figure}[h]
    \centering
    \subfloat[\textcolor{black}{Smooth spherical indenter, ${\bar{u}=0.0089\millm}$.}]{
    \includegraphics[width=0.45\linewidth]{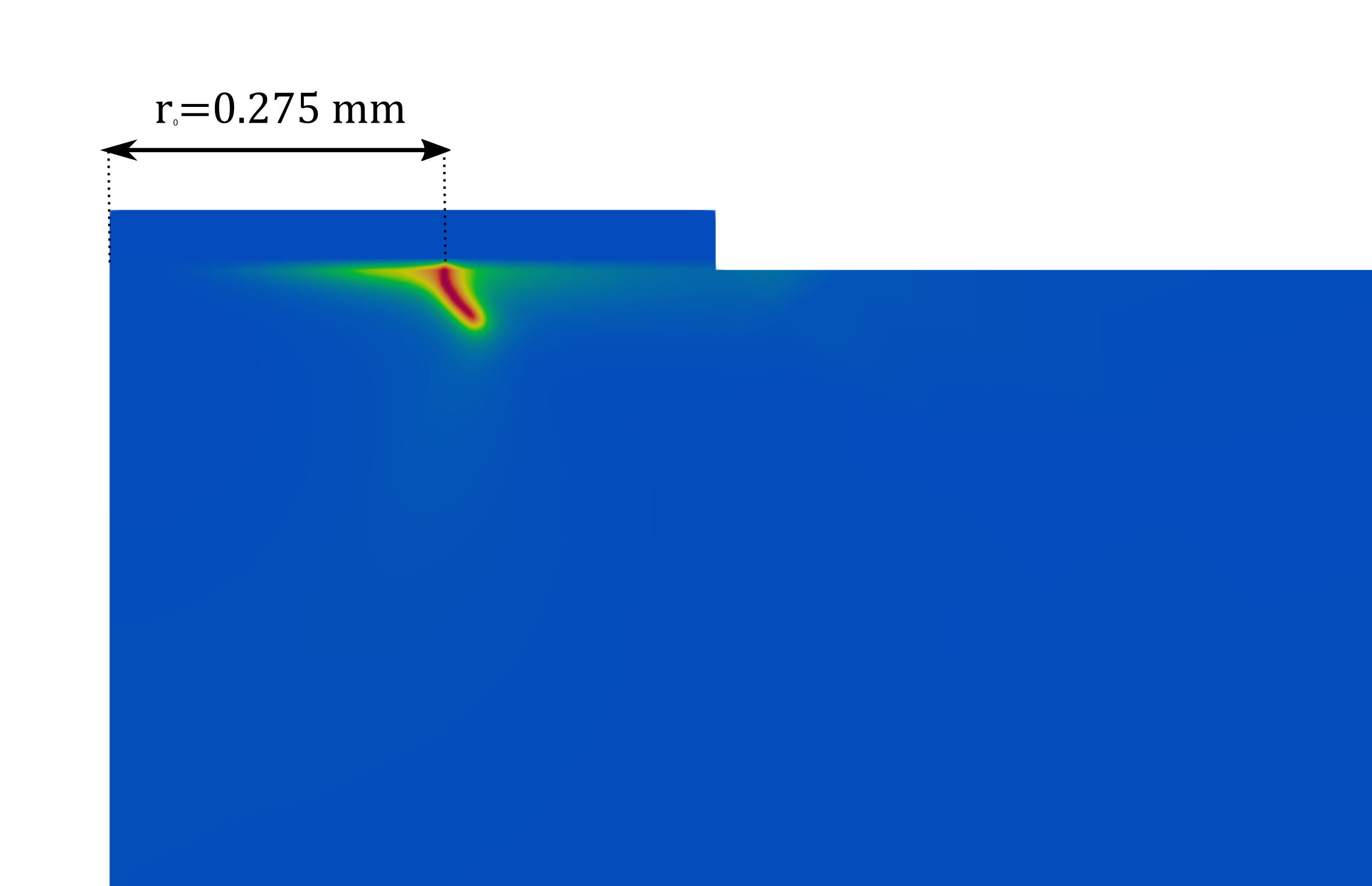}}\qquad
    \subfloat[\textcolor{black}{Rough spherical indenter with ${R_z=1.3\,\si{\mu m}}$, ${\bar{u}=0.0093\millm}$.} ]{
    \includegraphics[width=0.45\linewidth]{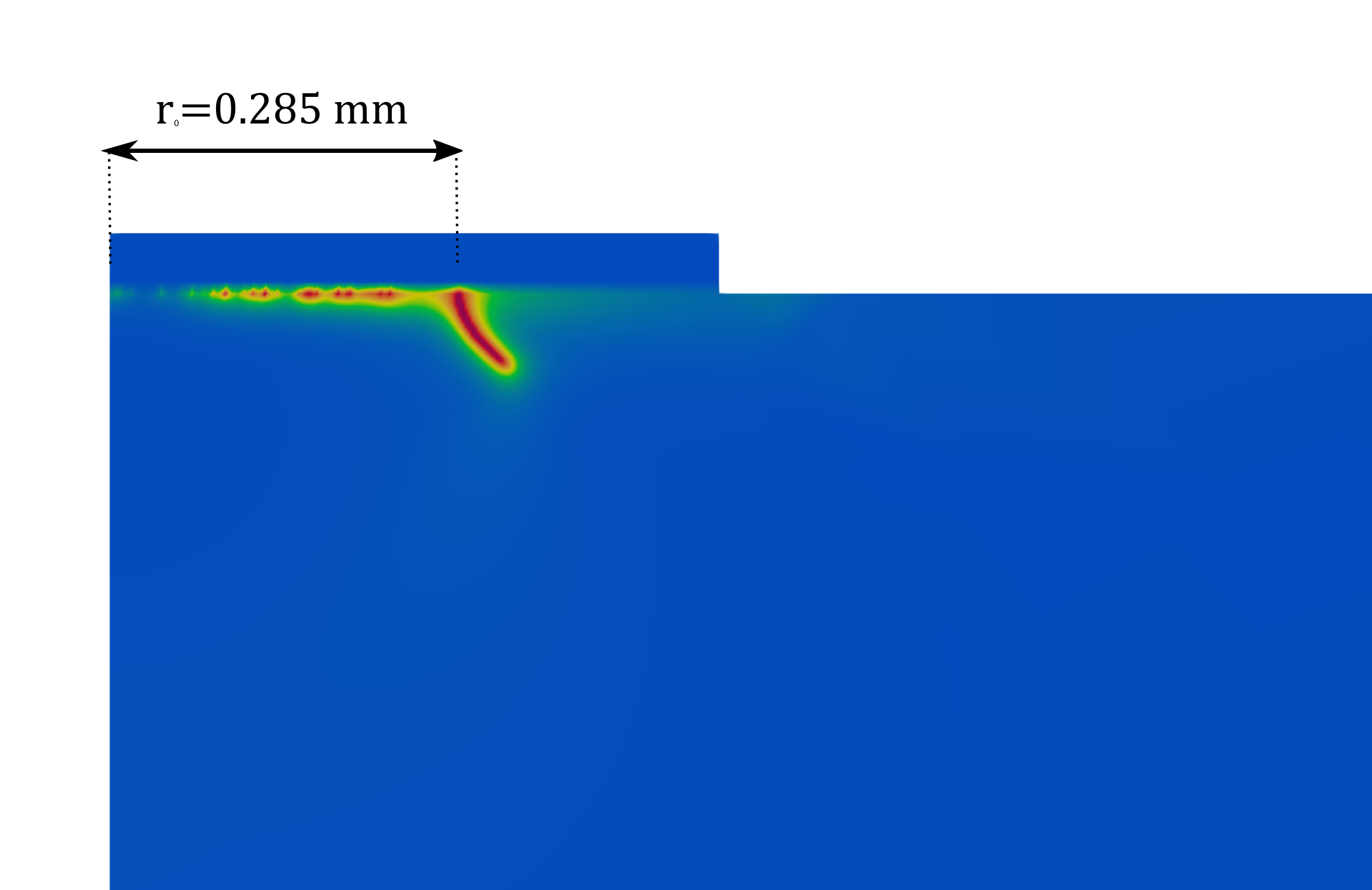}}\\
    \subfloat[\textcolor{black}{Rough spherical indenter with ${R_z=2.6\,\si{\mu m}}$, ${\bar{u}=0.012\millm}$.}]{\includegraphics[width=0.45\linewidth]{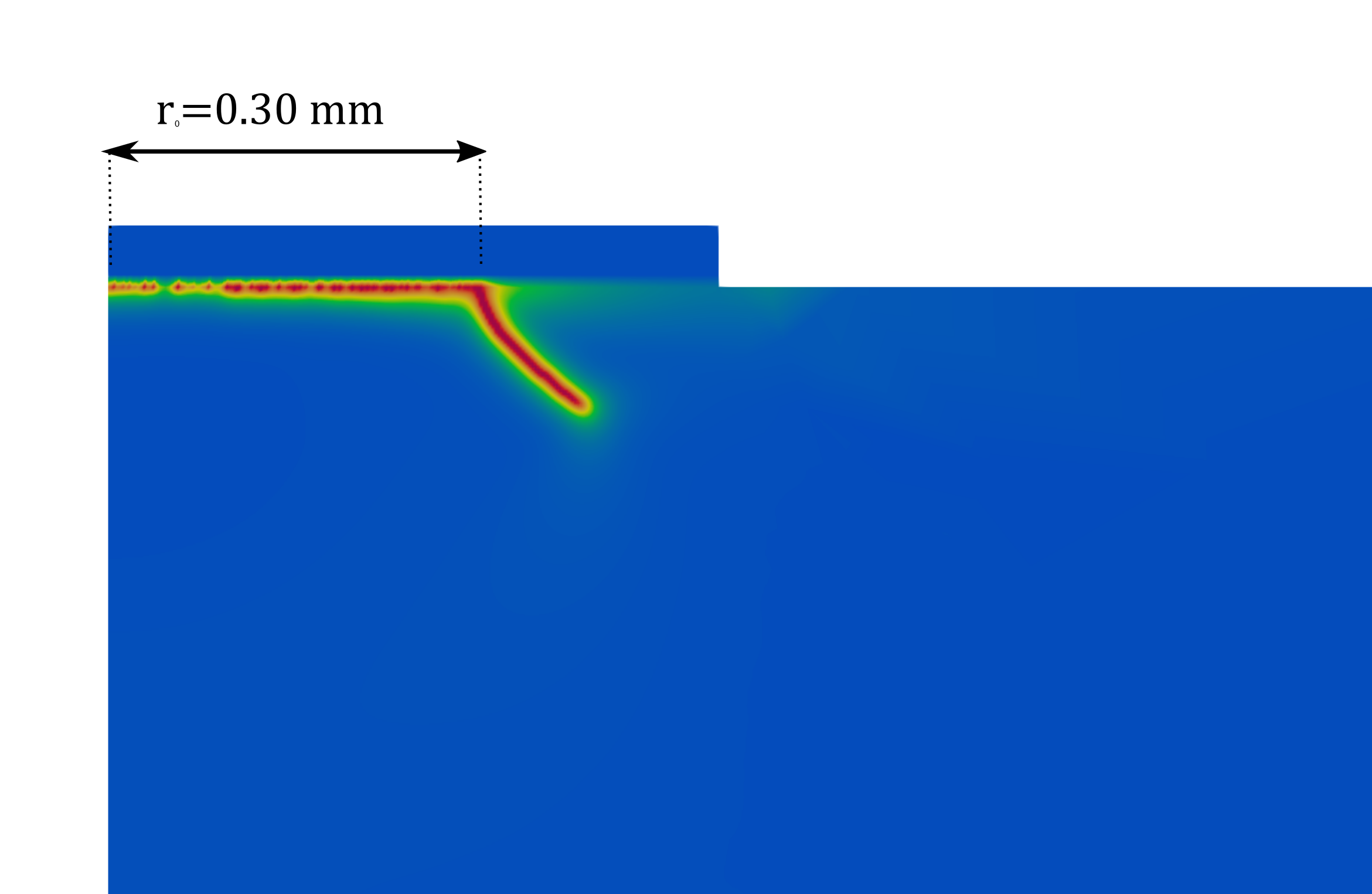}}\qquad
    \subfloat[\textcolor{black}{Rough spherical indenter with  ${R_z=5.2\,\si{\mu m}}$, ${\bar{u}=0.022\millm}$.}]{\includegraphics[width=0.45\linewidth]{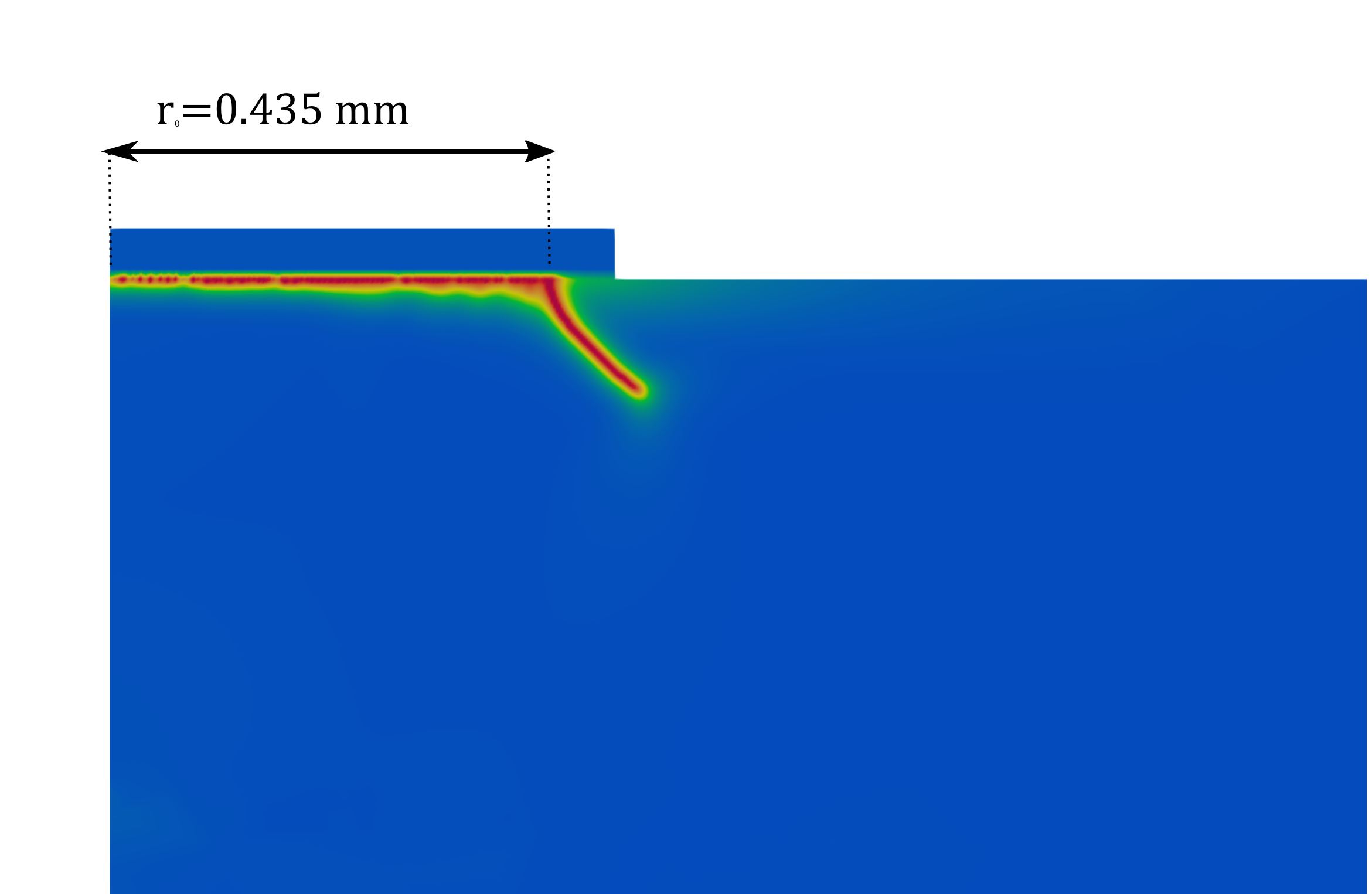}} \\
    \includegraphics[width=0.2\linewidth]{legend_horizontal.pdf}    
    \caption{\textcolor{black}{Radial position of the cone-shaped crack ($r_0$) on the surface of the glass substrate due to smooth and rough indenters with radius $R_s=5\millm$ and different amplitude of roughness, $R_z$.}}
    \label{fig:ring_crack_rad}
\end{figure}

\textcolor{black}{The rough profile presented in \fig\ref{fig:rough_prof} has been rescaled to simulate \textcolor{black}{three} different maximum peak-to-valley distances $R_z=1.3\,\si{\mu m}$, $R_z=2.6\,\si{\mu m}$, \textcolor{black}{and $R_z=5.2\mum$}.}
\textcolor{black}{The comparison with the corresponding smooth spherical indenter shows that the ring crack radius increases by amplifying roughness, see \fig\ref{fig:ring_crack_rad} where the predictions for the smooth case and the three rough profiles are shown. For each case, the contour plots are taken at the time step corresponding to the crack propagation.} 

\begin{figure}[h]
    \centering
    \includegraphics[width=0.8\linewidth]{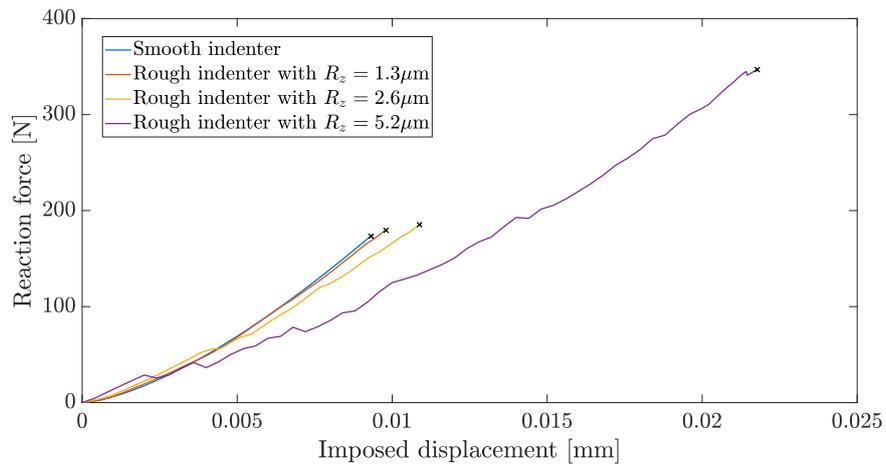}
    \caption{\textcolor{black}{Reaction force vs. far field imposed displacement for the smooth and rough indenters (radius of the sphere $R_s=5\millm$).}}
    \label{fig:reac}
\end{figure}

The presence of roughness also affects the critical load for the crack initiation, as highlighted in  
\fig\ref{fig:reac}. The result is in line with the experimental data in \citep{Conrad1979}, where an increase of the critical load in the case of abrasion of the glass substrate was reported. \textcolor{black}{In particular, for the indenter radius $R_s=5\millm$ analyzed here, \citep{Conrad1979} reported an increase of the critical load from $210 \si{N}$ to $300 \si{N}$ in case of abrasion with 600 SiC paper, while the simulations show an increase from $173 \si{N}$ to $346\si{N}$ for the rough profile with $R_z=5.2\mum$.}

\FloatBarrier

\section{Conclusion}
In this work, the simulation of cone-shaped cracks generated by spherical indentation tests has been successfully addressed by combining the MPJR interface finite elements to solve the non-conforming contact problem and the PF finite elements to predict nonlocal damage and fracture in the substrate. Coupling these two sources of nonlinearities, namely the contact mechanics problem and the fracture mechanics one, was one of the major challenges and contributions of the present work from the methodological point of view. In this regard, the MPJR interface finite elements have been demonstrated to be particularly efficient in treating complex interface profiles, as in the case of the simulated rough spherical indenters, where no previous solutions were obtained in the related literature. 

This computational framework allowed us to assess the indenter radius influence on the indentation cracking. The comparison with the benchmark experimental results from \cite{Conrad1979} demonstrated that the current variational formalism enables reproducing the critical load for crack initiation and the major features of the physical problem.

The influence of roughness at the contact interface has been investigated by considering three different spherical rough profiles with increasing roughness amplitude. The comparison with the smooth case showed an increase in the critical load for crack propagation and in the ring crack radius by increasing roughness, in agreement with the experimental trends available in the literature. \textcolor{black}{As an additional main insight into the problem, we obtained that the presence of roughness can induce higher stresses in the contact zone and therefore localized damage approaching $\phi=1$ in that area, which does not occur in case of smooth contacts. However, the origin of the propagating crack leading to cone fracture was always found outside the contact area.}

In terms of perspectives for future research, the present cases exploited the axial symmetry of the spherical indenter, which is computationally convenient. \textcolor{black}{The framework is already suitable for addressing 3D models where the entire rough surface can be incorporated into the FE simulations, even though it requires a high computational effort. A current line of research aims at developing a further HPC-enhanced framework to speed up the simulations.} Moreover, the developed framework opens new research directions in the field of indentation-induced fracture, also in the case of coated, FGM, or elasto-plastic substrates.

\FloatBarrier


\section*{CRediT authorship contribution statement}
\textbf{M. R. Marulli}: Methodology, Software, Investigation, Validation, Writing - Original Draft, Writing- Reviewing and Editing. \textbf{J. Bonari}: Software, Visualization, Writing- Reviewing and Editing. \textbf{J. Reinoso}: Conceptualization, Supervision, Writing- Reviewing and Editing. \textbf{M. Paggi}:  Conceptualization, Supervision, Original Draft, Writing- Reviewing and Editing. 

\section*{Acknowledgements}
The authors acknowledge the funding received from the European Union's Horizon 2020 research and innovation program under the Marie Skłodowska-Curie grant agreement No. 101086342 – Project DIAGONAL (Ductility and fracture toughness analysis of functionally graded materials; HORIZON-MSCA-2021-SE-01 action).




\section*{Appendix}

\textcolor{black}{To map the 3D nodal displacement vector $\bar{\bu}=(u_1,v_1,w_1,\phi_1,...,u_8,v_8,w_8,\phi_8)^\text{T}$ into a global field vector also including the nodal values of the phase-field damage variable,   
$\bar{\bu^e}=(u_1,v_1,w_1,\phi_1,...,u_8,v_8,w_8,\phi_8)^\text{T}$, the following matrix operator $\mathbf{P^{3D}}$ $(32\times 24)$ is introduced:
\begin{equation}
  \bar{\bu^e}=\mathbf{P^{3D}}\bar{\bu},  
\end{equation}
whose expression reads:
\begin{equation}
\mathbf{P^{3D}}=\left[
\begin{matrix}
\mathbf{I^{3\times 3}} & \mathbf{O^{3\times 3}} & \mathbf{O^{3\times 3}} & \mathbf{O^{3\times 3}} & \mathbf{O^{3\times 3}} & \mathbf{O^{3\times 3}} & \mathbf{O^{3\times 3}} & \mathbf{O^{3\times 3}} \\
\mathbf{O^{1\times 3}} & \mathbf{O^{1\times 3}} & \mathbf{O^{1\times 3}} & \mathbf{O^{1\times 3}} & \mathbf{O^{1\times 3}} & \mathbf{O^{1\times 3}} & \mathbf{O^{1\times 3}} & \mathbf{O^{1\times 3}} \\
\mathbf{I^{3\times 3}} & \mathbf{O^{3\times 3}} & \mathbf{O^{3\times 3}} & \mathbf{O^{3\times 3}} & \mathbf{O^{3\times 3}} & \mathbf{O^{3\times 3}} & \mathbf{O^{3\times 3}} & \mathbf{O^{3\times 3}} \\
\mathbf{O^{1\times 3}} & \mathbf{O^{1\times 3}} & \mathbf{O^{1\times 3}} & \mathbf{O^{1\times 3}} & \mathbf{O^{1\times 3}} & \mathbf{O^{1\times 3}} & \mathbf{O^{1\times 3}} & \mathbf{O^{1\times 3}} \\
\mathbf{I^{3\times 3}} & \mathbf{O^{3\times 3}} & \mathbf{O^{3\times 3}} & \mathbf{O^{3\times 3}} & \mathbf{O^{3\times 3}} & \mathbf{O^{3\times 3}} & \mathbf{O^{3\times 3}} & \mathbf{O^{3\times 3}} \\
\mathbf{O^{1\times 3}} & \mathbf{O^{1\times 3}} & \mathbf{O^{1\times 3}} & \mathbf{O^{1\times 3}} & \mathbf{O^{1\times 3}} & \mathbf{O^{1\times 3}} & \mathbf{O^{1\times 3}} & \mathbf{O^{1\times 3}} \\
\mathbf{I^{3\times 3}} & \mathbf{O^{3\times 3}} & \mathbf{O^{3\times 3}} & \mathbf{O^{3\times 3}} & \mathbf{O^{3\times 3}} & \mathbf{O^{3\times 3}} & \mathbf{O^{3\times 3}} & \mathbf{O^{3\times 3}} \\
\mathbf{O^{1\times 3}} & \mathbf{O^{1\times 3}} & \mathbf{O^{1\times 3}} & \mathbf{O^{1\times 3}} & \mathbf{O^{1\times 3}} & \mathbf{O^{1\times 3}} & \mathbf{O^{1\times 3}} & \mathbf{O^{1\times 3}} \\
\mathbf{I^{3\times 3}} & \mathbf{O^{3\times 3}} & \mathbf{O^{3\times 3}} & \mathbf{O^{3\times 3}} & \mathbf{O^{3\times 3}} & \mathbf{O^{3\times 3}} & \mathbf{O^{3\times 3}} & \mathbf{O^{3\times 3}} \\
\mathbf{O^{1\times 3}} & \mathbf{O^{1\times 3}} & \mathbf{O^{1\times 3}} & \mathbf{O^{1\times 3}} & \mathbf{O^{1\times 3}} & \mathbf{O^{1\times 3}} & \mathbf{O^{1\times 3}} & \mathbf{O^{1\times 3}} \\
\mathbf{I^{3\times 3}} & \mathbf{O^{3\times 3}} & \mathbf{O^{3\times 3}} & \mathbf{O^{3\times 3}} & \mathbf{O^{3\times 3}} & \mathbf{O^{3\times 3}} & \mathbf{O^{3\times 3}} & \mathbf{O^{3\times 3}} \\
\mathbf{O^{1\times 3}} & \mathbf{O^{1\times 3}} & \mathbf{O^{1\times 3}} & \mathbf{O^{1\times 3}} & \mathbf{O^{1\times 3}} & \mathbf{O^{1\times 3}} & \mathbf{O^{1\times 3}} & \mathbf{O^{1\times 3}} \\
\mathbf{I^{3\times 3}} & \mathbf{O^{3\times 3}} & \mathbf{O^{3\times 3}} & \mathbf{O^{3\times 3}} & \mathbf{O^{3\times 3}} & \mathbf{O^{3\times 3}} & \mathbf{O^{3\times 3}} & \mathbf{O^{3\times 3}} \\
\mathbf{O^{1\times 3}} & \mathbf{O^{1\times 3}} & \mathbf{O^{1\times 3}} & \mathbf{O^{1\times 3}} & \mathbf{O^{1\times 3}} & \mathbf{O^{1\times 3}} & \mathbf{O^{1\times 3}} & \mathbf{O^{1\times 3}} \\
\mathbf{I^{3\times 3}} & \mathbf{O^{3\times 3}} & \mathbf{O^{3\times 3}} & \mathbf{O^{3\times 3}} & \mathbf{O^{3\times 3}} & \mathbf{O^{3\times 3}} & \mathbf{O^{3\times 3}} & \mathbf{O^{3\times 3}} \\
\mathbf{O^{1\times 3}} & \mathbf{O^{1\times 3}} & \mathbf{O^{1\times 3}} & \mathbf{O^{1\times 3}} & \mathbf{O^{1\times 3}} & \mathbf{O^{1\times 3}} & \mathbf{O^{1\times 3}} & \mathbf{O^{1\times 3}} 
\end{matrix}    
\right],
\end{equation}
where $\mathbf{I^{3\times 3}}$ is a $(3\times 3)$ identify matrix, $\mathbf{O^{3\times 3}}$ is a $(3\times 3)$ null matrix, and $\mathbf{O^{1\times 3}}$ is a $(1\times 3)$ null matrix.}

\textcolor{black}{
For the 2D axisymmetric problem, the displacement field $\bar{\bu}=(u_1,v_1,...,u_4,v_4)^\text{T}$ is mapped onto the vector $\bar{\bu^e}=(u_1,v_1,\phi_1,...,u_4,v_4,\phi_4)^\text{T}$ with the following matrix operator $\mathbf{P^{2D}}$:
\begin{equation}
\mathbf{P^{2D}}=\left[
\begin{matrix}
     1 &    0  &   0  &   0  &   0  &   0  &   0  &   0\\
     0 &    1  &   0  &   0  &   0  &   0  &   0  &   0\\
     0 &    0  &   0  &   0  &   0  &   0  &   0  &   0\\
     0 &    0  &   1  &   0  &   0  &   0  &   0  &   0\\
     0 &    0  &   0  &   1  &   0  &   0  &   0  &   0\\
     0 &    0  &   0  &   0  &   0  &   0  &   0  &   0\\
     0 &    0  &   0  &   0  &   1  &   0  &   0  &   0\\
     0 &    0  &   0  &   0  &   0  &   1  &   0  &   0\\
     0 &    0  &   0  &   0  &   0  &   0  &   0  &   0\\
     0 &    0  &   0  &   0  &   0  &   0  &   1  &   0\\
     0 &    0  &   0  &   0  &   0  &   0  &   0  &   1\\
     0 &    0  &   0  &   0  &   0  &   0  &   0  &   0
\end{matrix}    
\right],
\end{equation}
The above operators have the fundamental property $\mathbf{P}^\text{T} \mathbf{P} =\mathbf{I}$ and the following inverse relation holds:  $\bar{\bu}=\mathbf{P}^\text{T}\bar{\bu^e}$.}

\bibliographystyle{agsm}

\end{document}